\documentclass[mathleft,fleqn,%
]{an}
\usepackage{graphicx}
\usepackage[varg]{txfonts}
\usepackage{natbib}
\bibpunct{(}{)}{;}{a}{}{,}

\usepackage{booktabs}	
\usepackage{subfigure}
\usepackage{url}
\usepackage[hidelinks=true, colorlinks=false, breaklinks=true, citecolor=black]{hyperref} 

\newcommand{\hi}{{H\sc{i}~}}
\newcommand{\kms}{$\mathrm{km\,s^{-1}}$}
\newcommand{\jb}{$\mathrm{Jy\,beam^{-1}}$}
\newcommand{\jkms}{$\mathrm{Jy\,km\,s^{-1}}$}

\begin{document}

\Pagespan{1}{}
\Yearpublication{2017}%
\Yearsubmission{2017}%
\Month{0}%
\Volume{999}%
\Issue{0}%
\DOI{To be added}%

\title{A new approach for short-spacing correction of radio interferometric data sets}

\author{S.\,Faridani\inst{1}\fnmsep\thanks{Corresponding author:
        {shahram.faridani@gmail.com}}
,  F.\,Bigiel\inst{2},  L.\,Fl\"oer\inst{1},  J.\,Kerp\inst{1}
\and S.\,Stanimirovi\'{c}\inst{3}
}
\titlerunning{A new approach for short-spacing correction in the image domain}
\authorrunning{Faridani et al.}
\institute{
Argelander-Institut f\"ur Astronomie (AIfA), Universit\"at Bonn, Auf dem H\"{u}gel 71, 53121 Bonn, Germany
\and 
Institut f\"ur theoretische Astrophysik, Zentrum f\"ur Astronomie der Universit\"at Heidelberg, Albert-Ueberle Str. 2, 69120 Heidelberg, Germany
\and 
Department of Astronomy, University of Wisconsin-Madison, 475 North Charter Street, Madison, WI 53706, USA}

\received{XXXX}
\accepted{XXXX}
\publonline{XXXX}


\abstract{%
The short-spacing problem describes the inherent inability of radio-interferometric arrays to measure the integrated flux and structure of diffuse emission associated with extended sources. New interferometric arrays, such as SKA, require solutions to efficiently combine interferometer and single-dish data.
\newline
We present a new and open source approach for merging single-dish and cleaned interferometric data sets requiring a minimum of data manipulation while offering a rigid flux determination and full high angular resolution. Our approach combines single-dish and cleaned interferometric data in the image domain. This  approach is tested for both Galactic and extragalactic \hi data sets. Furthermore, a quantitative comparison of our results to commonly used methods is provided. Additionally, for the interferometric data sets of NGC\,4214 and NGC\,5055, we study the impact of different imaging parameters as well as their influence on the combination for NGC\,4214.
\newline
The approach does not require the raw data (visibilities) or any additional special information such as antenna patterns. This is advantageous especially in the light of upcoming radio surveys with heterogeneous antenna designs.}

\maketitle

\section{Introduction}\label{sec:introduction}

Future radio-interferometer arrays will enable a new generation of \hi 21 cm line surveys for studying different scientific aspects of galaxy dynamics and evolution as well as the interstellar medium (ISM) \citep{2008ExA....22..151J, Dickey2013}. In the future, various surveys will be conducted with the Australia's Square Kilometer Array Pathfinder \citep[ASKAP, ][]{2009ASPC..407..446J} such as the GASKAP survey \citep{2013PASA...30....3D}, or with Meerkat \citep{2009pra..confE...7D}, the South African SKA pathfinder, such as the LADUMA survey \citep{2012IAUS..284..496H}. These instruments will not only measure the gas distribution of the Milky Way with high angular resolution, but also investigate the \hi content of galaxies in the local universe \citep{2010iska.meetE..43O, 2012MNRAS.426.3385D}. New technologies such as focal plane arrays increase the survey speed by more than an order of magnitude \citep{2011esci.conf...21N}. With these improvements, large-scale surveys at arcsecond angular resolution become feasible.

For these new facilities one of the major issues will be handling the huge amount of data \citep{2011esci.conf...21N, 2013PASA...30....3D}. For these surveys the long term storage of the raw data (visibilities) is actually not planned and most likely not feasible. Automated on-the-fly data reduction pipelines and parameterized source finding algorithms will be used \citep{2012PASA...29..318P, 2012MNRAS.421.3242W, 2015MNRAS.448.1922S} to extract science-ready data from these observations. 

All these new facilities are interferometric arrays, which are, by design, subject to the so-called short-spacing problem \citep[][]{1985A&A...143..307B}. The lack of very short baselines leads to insensitivity to emission from large angular scales. This is particularly an issue when observing the neutral Galactic ISM or diffuse HI halos around galaxies \citep{1999MNRAS.302..417S}. In this respect, single-dish telescopes will still be important to study Galactic \hi emission or the diffuse ISM in nearby galaxies. Single dishes provide the missing-spacing information which can be added to an interferometric observation. This process is called the short-spacing correction \citep[SSC, ][]{2002ASPC..278..375S}. The aim of this procedure is to recover the integrated flux density and diffuse emission, as measured with a single dish, while preserving the high angular resolution of an interferometer.

Interferometry is an important observational approach for wavelengths from sub-millimeter, millimeter to the cm regime and beyond. SSC is important and desirable especially for instruments such as ALMA and NOEMA \citep[e.g., ][]{2013ApJ...779...42S}, where the combination of interferometric and single-dish data sets is indispensable and constantly used. 

However, the combination of single-dish and interferometric data sets is non-trivial. Typically, either individual SSC implementations are developed for a specific purpose or data set, or require the user to adjust various parameters \citep[e.g., ][]{2006AJ....132.1158S, 2016arXiv160202115B}. This situation motivated us to perform a detailed investigation of standard SSC schemes and eventually led to the development of a new approach, which operates in the image domain. Moreover, this new approach is ideally suited also for online use on the large data sets from future facilities.

In addition, methods using the Fast Fourier Transformation (FFT) inherently adopt periodicity of the signal. This may cause artifacts produced by structures close to the edge of or even beyond the primary beam of the radio interferometer. This is again a common case for observations of Galactic extended structures, where a significant amount of bright emission is located at the borders of each map \citep{2016A&A...592A.142R}.

We also investigate differences between existing methods for the SSC from the perspective of the observer. Analyzing real observations rather than simulations allows to test our approach by adopting inherently realistic conditions (artifacts, calibration offsets, radio frequency interferences (RFI), unstable baselines, etc.). Especially, we use \hi observations of the Small Magellanic Cloud \citep[SMC, ][]{1997MNRAS.289..225S, 1999MNRAS.302..417S} and different interferometric \hi data cubes of NGC\,4214 and NGC\,5055 from The \hi Nearby Galaxy Survey \citep[THINGS, ][]{2008AJ....136.2563W}. These different objects with large and variable angular extents on the sky provide ideal test cases for our method.

Additionally, we investigate the impact of two different imaging parameters, i.e., weighting scheme and pixel size on the result of the interferometric data and combination, respectively. We investigate the changes of different characteristics in the interferometric map, in particular the flux density, as a function of the aforementioned imaging parameters.

These parameters are of great importance for the combination, since the characteristics of each interferomteric map affect the result of combination significantly. Furthermore, long term storage of interferomteric raw data is not feasible for the upcoming interferometric facilities. Therefore, the data reduction process can not be repeated arbitrarily.

The structure of the paper is as follows: Section\,\ref{sec:synthesis_im} provides a brief introduction to the principle of synthesis imaging, the short-spacing problem, and the princible solution to perform the SSC. Section\,\ref{sec:data} presents the details of the SMC, NGC\,4214, and NGC\,5055 data. Section\,\ref{sec:pipeline} describes our approach for performing the SSC in the image domain. Section \ref{sec:evaluation} presents the evaluation of our approach. Section\,\ref{sec:comp_method} comprises a comparison of the results of combination for the SMC data sets using three different methods. Section\,\ref{sec:syn_image_par} discusses the interferometric imaging parameters weighting scheme and pixel size and Section\,\ref{sec:impac_imag_par} discusses their impact on the flux distribution of the resulting synthesized image and the SSC. Section\,\ref{sec:summary} summarizes our results and provides an outlook regarding possible future work.

\section{Synthesis imaging and missing spacings}\label{sec:synthesis_im}

The smallest angular scale that can be resolved decreases with increasing frequency and telescope diameter. In synthesis imaging, signals from a large number of medium sized telescopes are combined. In principle, the largest separation between the array dishes (longest baseline) determines the best achievable angular resolution.

The synthesized image presents a best model of the true sky intensity distribution $I^{(\nu)}(l,m)$ of a source as a function of direction cosines $l,m$. The final image is reconstructed from a non-uniformly sampled visibility function $V^{(\nu)}(u,v)$ via the Fourier transformation, where the observation is conducted for a specific frequency $\nu$, i.e., line observation or a finite range $\nu$ in case of continuum observations. Equation\,\ref{eq:vis} describes the relationship between the visibility $V^{(\nu)}(u,v)$ and the brightness distribution $I^{(\nu)}(l,m)$ 

\begin{equation}\label{eq:vis}
V(u,v) = \iint I(l,m)A(l,m)e^{-2\pi i(ul+vm)}\,dldm,
\end{equation}

where $A(l,m)$ is the primary beam of a single antenna of the array at a certain frequency $\nu$. Visibilities $V(u_{ij}(t), v_{ij}(t))$ are measured in the $(u,v)$-domain, where each sample is the cross correlation of incoming signals for an antenna pair $(i,j)$. $(u_{ij}(t), v_{ij}(t))$ is referred to as a baseline $\overrightarrow{b}$. 

Henceforward, we neglect the degradation of the visibilities due to integration time and finite bandwidth \citep[and references therein]{1999ASPC..180...11T}.

An interferometer samples the visibility space at discrete points given by the array properties. In this case, one writes:

\begin{eqnarray}\label{eq:true_vis}
V^{(\nu)}_\mathrm{\,obs}(u, v) = V^{(\nu)}_\mathrm{\,true}(u, v) \cdot S^{(\nu)}(u, v),
\end{eqnarray}

with the sampling function $S^{(\nu)}(u,v)$: 

\begin{eqnarray}
S^{(\nu)}(u,v)=\left\{
\begin{array}{cl}
	1, & \mbox{if } (u,v) \in \mbox{observations} \\ 
	0, & \mbox{otherwise}  
\end{array}\right.
\end{eqnarray}

Using the inverse Fourier transformation and the convolution theorem, one retrieves the \textit{dirty image} ($I^{D}(\xi, \eta)$) and \textit{dirty beam}. The latter is the inverse Fourier transformation of the sampling function $S_{(\nu)}(u, v)$. 

\begin{eqnarray}\label{eq:dirty_im}
\mathfrak{F}^{-1}(V^{(\nu)}_\mathrm{obs}(u,v)) & = & \mathfrak{F}^{-1}(V^{(\nu)}_\mathrm{true}(u,v) \cdot S^{(\nu)}(u,v)) \\
 I^{D}(\xi, \eta) & = & \mathfrak{F}^{-1}(V^{(\nu)}_\mathrm{true}(u,v)) \ast \mathfrak{F}^{-1}(S^{(\nu)}(u,v)) \nonumber.
\end{eqnarray}

$V^{(\nu)}_\mathrm{true}(u,v)$ and $V^{(\nu)}_\mathrm{obs}(u,v)$ are true and observed visibilities. The ``$\ast$'' symbol denotes the convolution of the two inverse Fourier transformations.

Measuring as many data points as possible in the $(u,v)$-plane is essential for an interferometric observation. However, one cannot sample the entire $(u,v)$-plane. Each missing baseline means that certain spatial frequencies are not measured. Thus, the integral is not uniquely solvable and can only be determined approximately. However, in practice it is impossible to position antennas at arbitrarily many locations. Aperture synthesis makes use of the Earth's rotation to increase the $(u,v)$-coverage. Mosaicing is another approach to improve the $(u,v)$-sampling, where a measurement consists of a concatenation of different pointings \citep{1999ASPC..180..401H, 2002ASPC..278..375S}.

Before the interferometric data can be used for scientific purposes, \textit{deconvolution} is necessary. Deconvolution tries to reconstruct the true brightness distribution from the limited sample of visibilities. The most common deconvolution technique is the CLEAN algorithm as introduced by \citet{1974A&A....33..289H} and its variants \citep[e.g., ][]{1980A&A....89..377C}. 

\subsection{Short-Spacing Problem (SSP)}\label{subsec:ssp}

For any interferometer the central region of the $(u, v)$-plane is never sampled ($u=v=0$). This is due to the physical size of the dishes and their minimal separation, which is referred to as the shortest baseline. The incompleteness of the $(u, v)$-coverage at low spatial frequencies, known as the short-spacing problem (SSP), leads to an insensitivity of interferometers towards emission on large angular scales. $u=v=0$ contains the total power information. The total flux density for a source is then given by:

\begin{eqnarray}
V(0,0) = \iint I(l,m)\,dldm = \int I \,d\Omega = S_\mathrm{tot}.
\end{eqnarray}
 
Note that the integrated flux density in the dirty image is zero \citep{synthesis_nrao}. By \textit{cleaning}, part of the integrated flux density can be reconstructed. The effect of short spacings is negligible for objects that are small in comparison to the extent of the primary beam. For Galactic objects and nearby galaxies however, which are large extended structures, the lack of sensitivity towards low spatial frequencies is a severe shortcoming. One specific example are the diffuse, low-column density, extended \hi disks around galaxies \citep[e.g., ][]{2005ApJ...619L..79T, 2010ApJ...720L..31B, 2010AJ....140.1194B}.

The interferometric observations of these objects suffer from the so-called negative bowls. These denote an image degradation that arises due to the lack of information on emission from large angular scale structures during the imaging process.  

To overcome the missing-spacing problem, the data from a single-dish telescope is required to fill the gap because these only measure total power.

\begin{figure}[!tb]
\centering
\includegraphics[width=0.95\columnwidth]{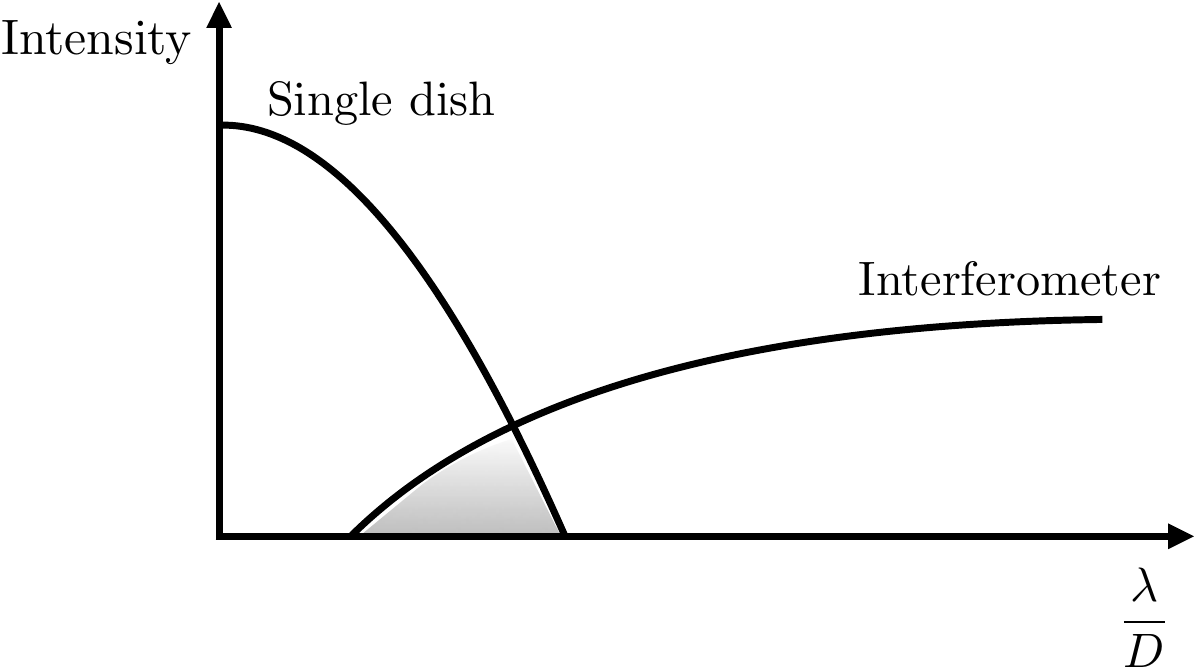}
\caption{Schematic view of the overlap region between single dish and interferometer. The overlap region corresponds to spatial scales towards which both instruments are sensitive.}
\label{fig:overlap}
\end{figure}

For the combination the $u,\,v$ overlap region of both is of great importance \citep{2002ASPC..278..375S}. Here, both instruments are sensitive (Fig.\,\ref{fig:overlap}) for the object's structure.

The most common techniques for adding missing spacings are part of one of the major astronomical data reduction packages (e.g. \texttt{imerg} in AIPS \citep{1999ascl.soft11003A}, \texttt{feather} in CASA \citep{2011ascl.soft07013I} and \texttt{immerge} in MIRIAD \citep{2011ascl.soft06007S}), where the combination occurs in the Fourier domain using single-dish and deconvolved interferometric data. The combination can also be performed prior to deconvolution. In this case a proper combined beam is necessary as shown by \citet{2002ASPC..278..375S}. 

Next, we introduce the data sets that are used to compare our combination method in the image domain to these established techniques.

\section{Observations and data}\label{sec:data}

In this section, we present the Small Magellanic Cloud (SMC), NGC\,4214 and NGC\,5055 \hi data sets. The SMC data sets presented stem from \citet{1999MNRAS.302..417S}. They used \hi observations obtained with the 64m Parkes telescope and the Australia Telescope Compact Array (ATCA)\,\citep{2002ASPC..278..375S}. We use the SMC data sets to evaluate the performance of our approach. Furthermore, we use NRAO VLA observations of NGC\,4214 \citep{1998PASA...15..157M, 2012IAUS..284..156H, 2013MNRAS.433.1276S, 2013ApJ...767...51A} and NGC\,5055 \citep{2006A&A...447...49B, 2012AAS...21934625P} obtained as part of the THINGS survey \citep{2008AJ....136.2563W} in order to demonstrate the impact of the imaging parameters (pixel size and weighting scheme) on the interferometric data. 

We do not use the SMC data sets to study the impacts of different imaging parameters, since the data set is an interferometric mosaic consists of 320 pointings \citep{1997MNRAS.289..225S, Stanimirovic_phd_1999}. The $(u,v)$-coverage of this observation is very complex and therefore, inappropriate for the purpose of our study. Whereas, the interferometric observations of NGC\,4214 and NGC\,5055 are single pointings. 

\subsection{The SMC observations and data}\label{sec:smc_data}

The Small Magellanic Cloud (SMC) is a nearby dwarf galaxy located at a distance of approximately 60 kpc \citep{2014ApJ...780...59G}. The measured \hi mass for the SMC varies between $3.4\times 10^{8} \leq M_{\odot} \leq 5.5 \times 10^{8}$ \citep{1982A&AS...48...71B, 1999MNRAS.302..417S, 2003ApJ...586..170P, 2005A&A...432...45B}. The variations in the measured \hi masses are probably caused by the different field of views of various observations. The galaxy reveals a complex morphology with a non-symmetric shape (Fig.\,\ref{fig:smc_data}). Various studies show that the galaxy has a strong filamentary structure with small, compact clumps embedded in a considerable amount of diffuse gas \citep{1997MNRAS.289..225S, 1999MNRAS.302..417S}. This is obvious in the interferometric and single-dish observations of the galaxy. The presence of both warm and cold components in combination with the nearby location makes it an ideal test object for various SSC methods.

While the single-dish observation reveals the non-symmetric shape of the SMC, the interferometric observation show a wealth of small-scale structures. Both observations cover an area of approximately 20 $\mathrm{degrees^{2}}$\,\citep{1997MNRAS.289..225S, Stanimirovic_phd_1999}. The angular resolution of the single-dish data is 18.8', the corresponding value for the interferometric data is 98''. The spectral resolution in the interferomteric and regridded single-dish data cubes is 1.65\,\kms with heliocentric velocities $88 \leq v_\mathrm{\,helio} \leq 216$\,\kms. The measured rms noise level in the low- and high-resolution data sets are $\approx 145$\,m\jb and $\approx 18$\,m\jb. 

The quality of both single-dish and interferometric data is of great importance for the combination. Therefore, \citet{1997MNRAS.289..225S, 1999MNRAS.302..417S} performed different calibration and data editing measures (e.g., removing solar interference etc.) to obtain the best possible image for both data sets.

Both the Parkes and ATCA images were tapered by multiplying by a function which smoothly decreased the image intensities to zero near the edges. This is important, since the Parkes image in particular has non-zero emission observed all over the map. The sharp edges produce strong horizontal and vertical ringing (spikes) in the center of the Fourier plane \citep{2002A&A...393..749M} after Fourier transforming \citep{Stanimirovic_phd_1999}.

\subsection{NGC\,4214 and NGC\,5055 observations and Data}\label{subsec:ngc4214_ngc5055_data}

In the following, we present different data sets of NGC\,4214 and NGC\,5055 observations. The reason for this selection is that these galaxies are located at different distances and they have substantial differences in their morphology, extent, and physical parameters \citep{2009AJ....137.4670L}. Table\,\ref{tb:n4214_n5055_obs_par} presents the physical and observational parameters of these two galaxies. They also differ in the amount of faint diffuse gas present in and around the galaxy \citep{2008AJ....136.2563W}. Furthermore, these galaxies are well separated from Galactic emission in velocity \citep[their Fig.\,1]{2008AJ....136.2563W}.

\begin{table}[hptb]
\centering
\resizebox{\linewidth}{!}{%
\begin{tabular}{l c c c c}
\toprule
Galaxy			& Type & Dist. & $\mathrm{v_{\,sys}}$ & $r_{\,25}^{\,a}$ \\
			&      & [Mpc] & [\kms] & [arcmin] \\
\midrule
NGC\,4214		& irr. dwarf & 2.9 & 291 & 3.4 \\
NGC\,5055		& sbc & 10.1 & 484 & 5.9 \\
\bottomrule
\end{tabular}}
\caption{Characteristics of NGC\,4214 and NGC\,5055. Note that $r_{\,25}$ is the radius of the B-band 25\,mag $\mathrm{arcsec^2}$ isophote \citep{2009AJ....137.4670L}.}
\label{tb:n4214_n5055_obs_par}
\end{table}

For each of these galaxies, four different interferometric data cubes have been imaged. New data sets have been produced based on the VLA raw data from the THINGS observations. The data reduction has been performed using the THINGS pipeline (Bigiel, priv. comm.). The details of the imaging process are presented by \citet{2008AJ....136.2563W}. Note that all the presented data in this work are corrected for the primary beam efficiency.

These four data cubes differ in the applied weighting scheme and pixel size. For consistency, the pixel sizes and number of pixels along R.A. and Dec. axes have been chosen such that all the data sets have the same FoV ($\approx 0.4 \, \mathrm{degree}$). 

For each galaxy, four data sets are imaged using robust parameters 5 and 0.5. While 5 is nearly pure natural weighting and achieves data sets with higher sensitivity, the latter robust parameter produces data sets with higher angular resolution. Henceforward, we refer to these data sets as NA for natural (robust parameter 5) and UN for uniform (robust parameter 0.5) weighted, respectively. Note that the used parameters are from AIPS and differ for other astronomical frameworks.

For each pair of data sets with natural and robust weighting, two different pixel sizes of 1.5'' and 3'' are chosen. Table\,\ref{tb:n4214_5055_phys} summarizes the important characteristics of both NGC\,4214 and NGC\,5055 interferometric data sets. The last column shows the measured total flux in the unmasked velocity-integrated intensity maps. Note that both angular resolutions of data sets as well as the amount of measured integrated flux densities change for different data sets depending on the applied weighting scheme and chosen pixel size. 

\begin{table}[hbpt]
\resizebox{\linewidth}{!}{%
\centering
\begin{tabular}{c c c c c c c c c}
\toprule
NGC\,4214               & Pixel     & Weight.  &  Beam         & Tot.\,flux        \\
                        & [arcsec]       &            & [arcsec]            & [\jkms]  \\
\midrule
$512 \times 512$        & 3              & NA         & $18.7 \times 19.8$  & 116.1             \\
$512 \times 512$        & 3              & UN         & $8.7 \times 8.8$    & -0.57             \\
$1024 \times 1024$      & 1.5            & NA         & $13.8 \times 14.6$  & 106.1             \\
$1024 \times 1024$      & 1.5            & UN         & $6.3 \times 7.4$    & -19.4             \\
\midrule
NGC\,5055\\
\midrule           
$512 \times 512$        & 3             & NA          & $10.4 \times 12.6$   & 263.5            \\
$512 \times 512$        & 3             & UN          & $7.6 \times 8.0$     & 110.4            \\
$1024 \times 1024$      & 1.5           & NA          & $8.6 \times 10.1$    & 255.9            \\
$1024 \times 1024$      & 1.5           & UN          & $5.3 \times 5.8$     & 81.1             \\        
\bottomrule
\end{tabular}}
\caption{The natural (RA) and uniform (UN) weighted data cubes of NGC\,4214 and NGC\,5055. In the process of imaging, for each data set two different pixel sizes have been chosen. However, the data sets have the same FoV. Beam size corresponds to the FWHM of the beam minor and major axis in each data cube. The measured total flux for each data cube is presented.}
\label{tb:n4214_5055_phys}
\end{table}

\section{The new SSC approach applied to the SMC data}\label{sec:pipeline}
We present our new and open source approach for combing single-dish and cleaned interferometric data sets in the image domain \footnote{The code is publicly available. \newline \url{https://bitbucket.org/snippets/faridani/pRX6r}}. 

In the first step both input FITS\footnote{Flexible Image Transform System (FITS) is a standardized file format commonly used for storing astronomical data.} files (low- and high-resolution data sets) are imported and the angular resolutions of the single-dish and interferometric data are retrieved from the corresponding FITS headers. Additionally, we check if both data sets have compatible brightness units. Since, low- and high-resolution data sets have different coordinate systems and projections, regridding is required. Regridding is the process of interpolation from a specific coordinate grid to a different one. The chosen default regridding scheme is linear interpolation which conserves surface brightness/intensity. This is a crucial factor for the combination. Linear interpolation is prone to block-like artifacts. The significance of the artifacts is higher when the difference in the grid resolution of both low- and high-resolution is large. The need for interpolation can be circumvented if an appropriate pixel grid is chosen during the single-dish data reduction process (due to the lower resolution of the single-dish data these are often on a coarser grid, i.e., the angular size of each pixel is larger). 

Note that if the intensities are in units of Jy/beam, the units are different for the interferomteric and the single-dish data sets ($\mathrm{Jy}/\mathrm{beam_{\,int}}$ and $\mathrm{Jy}/\mathrm{beam_{\,sd}}$, respectively). This is the reason why Eq.\,\ref{eq:ssc_shf} contains the correction factor $\alpha$, which is the ratio of the interferometric and the single-dish beam areas: $\alpha = \mathrm{beam\,area}_\mathrm{\,int}/\,\mathrm{beam\,area}_\mathrm{\,sd}$.

Furthermore, It is important to note that for data sets in units of K, Jy/pixel, Jy/arcseconds, etc., the factor $\alpha$ is not necessary and must be omitted from the Eq.\,\ref{eq:ssc_shf}. Our pipeline recognises and treats all these units appropriately.

For the determination of missing flux, the interferometric data are convolved with a two dimensional, normalized Gaussian kernel such that the angular resolution of the convolved interferometric data matches the single dish data. The difference between the convolved interferometric and the regridded single-dish data cube is proportional to the missing flux, i.e., the missing information that the interferometer lacks.

The additional flux is added to the interferometric data set and the combined data set is exported. Mathematically, this is written as follows:

\begin{eqnarray}\label{eq:ssc_shf}
I_\mathrm{\,missing} & = & I_\mathrm{\,sd}^\mathrm{\,reg} - I_\mathrm{\,int}^\mathrm{\,conv}\\
I_\mathrm{\,comb} & = & I_\mathrm{\,int} + \alpha \cdot I_\mathrm{\,missing} \nonumber.
\end{eqnarray}

$I_\mathrm{\,missing}$ is the missing flux only observed by the single dish. $I_\mathrm{\,sd}^\mathrm{\,reg}$ is the regridded single-dish data set, $I_\mathrm{\,int}^\mathrm{\,conv}$ the convolved interferometric data set as described before. The combination $I_\mathrm{\,comb}$ is the result of summation of the interferometric data set $I_\mathrm{\,int}$ and the missing flux multiplied by $\alpha$.

\begin{figure*}[!tb]
\centering
\includegraphics[width=0.89\textwidth]{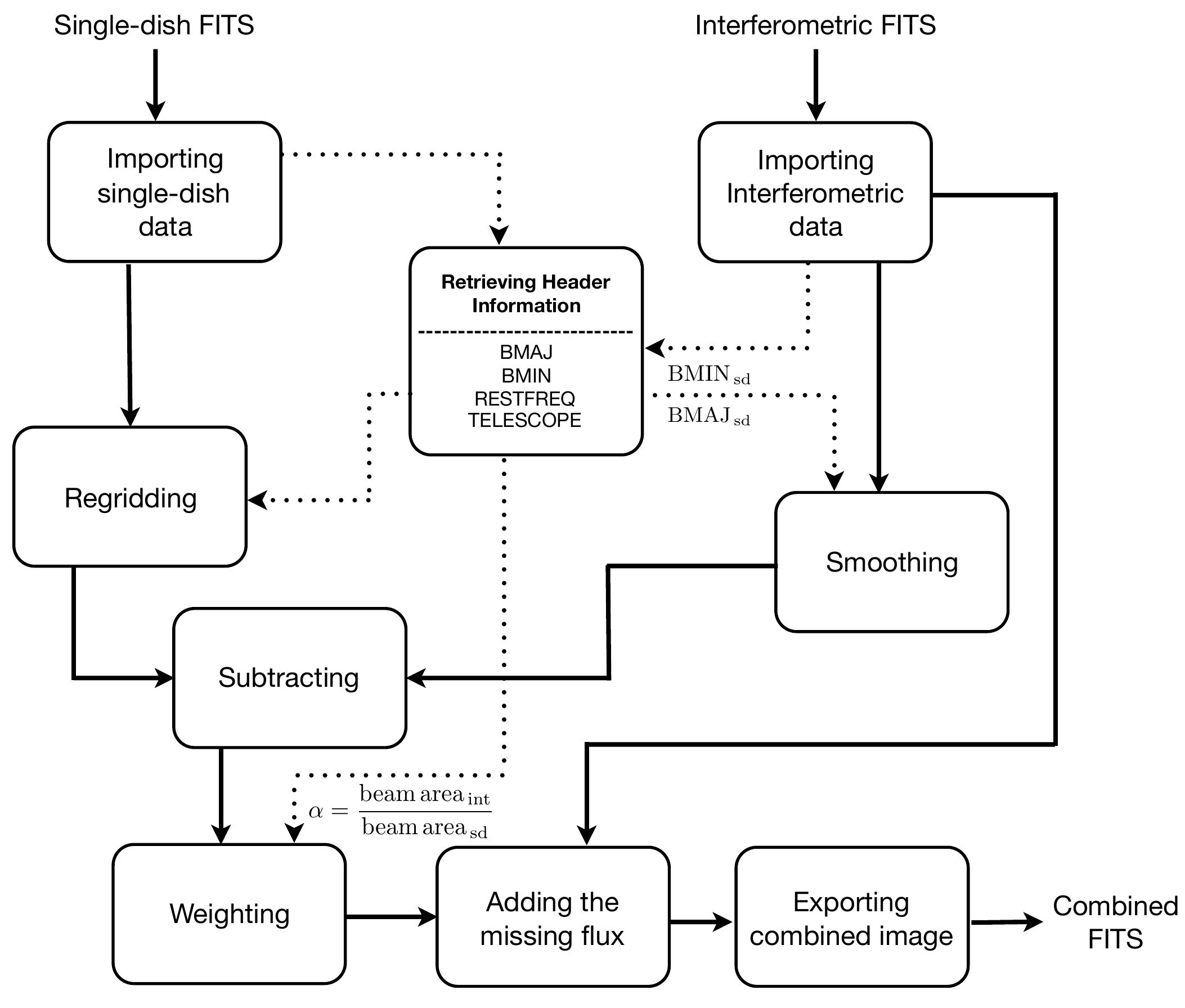}
\caption[Data flow of the short-spacing \textbf{approach}]{Data flow scheme of the developed short-spacing method in the image domain. The dotted lines represent the metadata retrieved from the header. The solid lines present the data flow. The inputs are the low- (single-dish) and high-resolution (interferometric) data cubes. The output is the combined data cube.}
\label{fig:pipeline}
\end{figure*}

Figure \ref{fig:pipeline} shows the data flow of the developed short-spacing approach, where the solid lines show the data flow and the dotted lines the retrieved information from the header. 

Our code is written in Python and makes use of existing CASA tasks and image tools \citep{2011ascl.soft07013I}.

\section{Evaluation of the SSC method for SMC data}\label{sec:evaluation}

The measured total flux densities in the low- and high-resolution data cubes are $4.5 \times 10^{5}$\,\jkms\, and $1.4 \times 10^{5}$\,\jkms, respectively. Hence, the interferometer measures less flux than the single dish. The corresponding value in the combined data cube is $4.5 \times 10^{5}$\,\jkms. The interferometer only receives about $30\%$ of the total flux. The recovered angular resolution in the combined data is 98''. 

The mean rms noise level in the combined data cube is about 20 m\jb, which is slightly higher than the corresponding value in the high-resolution data set. It is however considerably lower than that of the value of the low-resolution data cube. Nevertheless, it is important to recall that the noise in both, the interferometric and the combined map, is a strong function of the considered angular scales. The noise level is directly related to the sampling of the $(u,v)$-plane.

\begin{figure*}[!tb]
	\centering
	\subfigure[SMC - ATCA]{\includegraphics[width=0.41\textwidth]{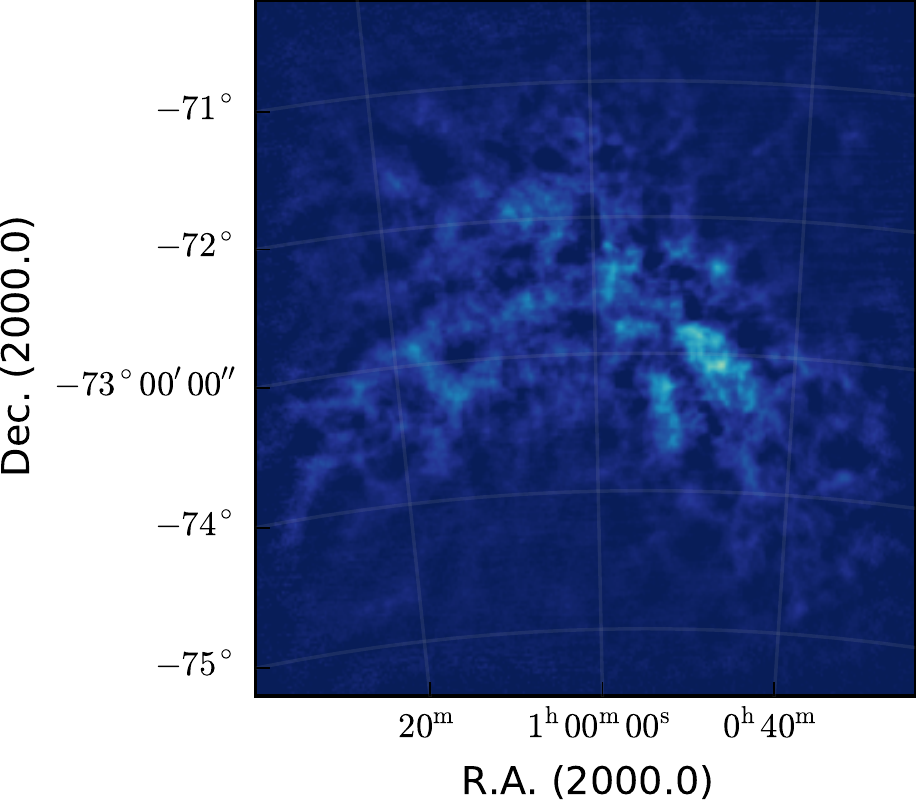}}
	\subfigure[SMC - Combined]{\includegraphics[width=0.54\textwidth]{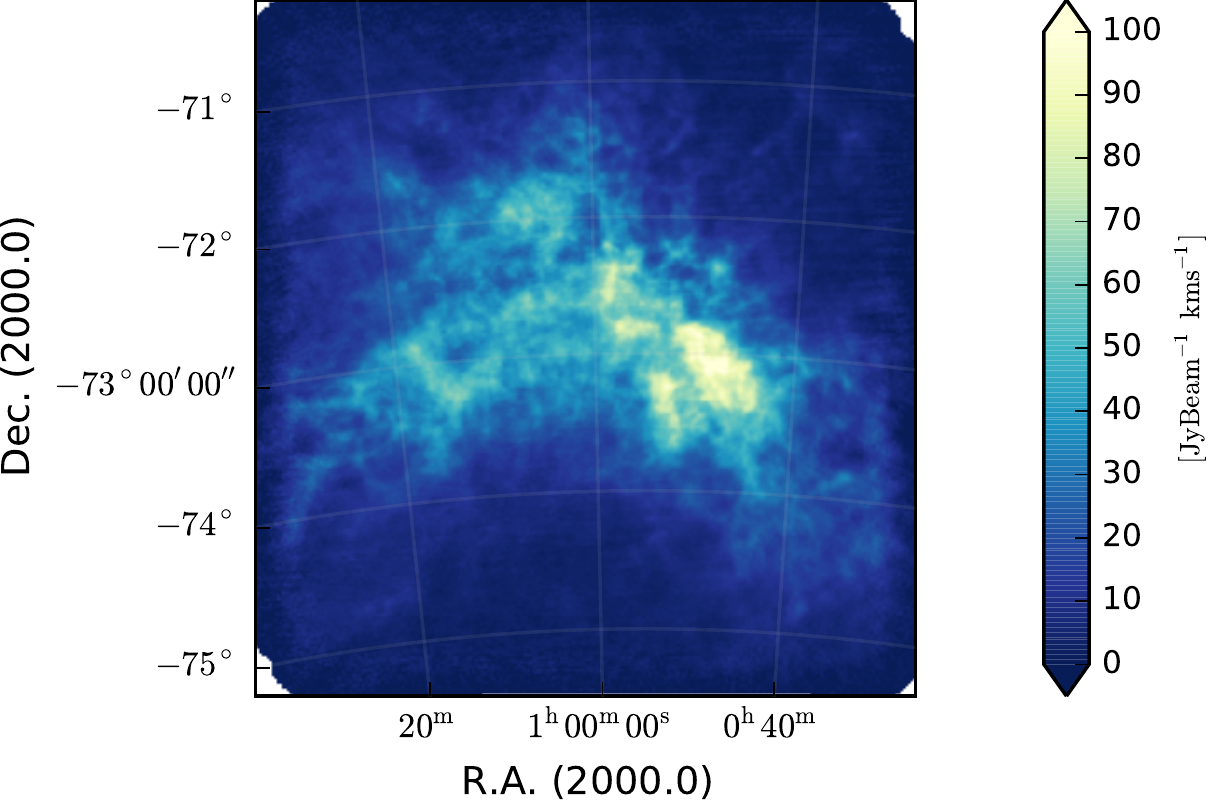}}
	\subfigure[SMC - Relative contribution]{\includegraphics[width=0.53\textwidth]{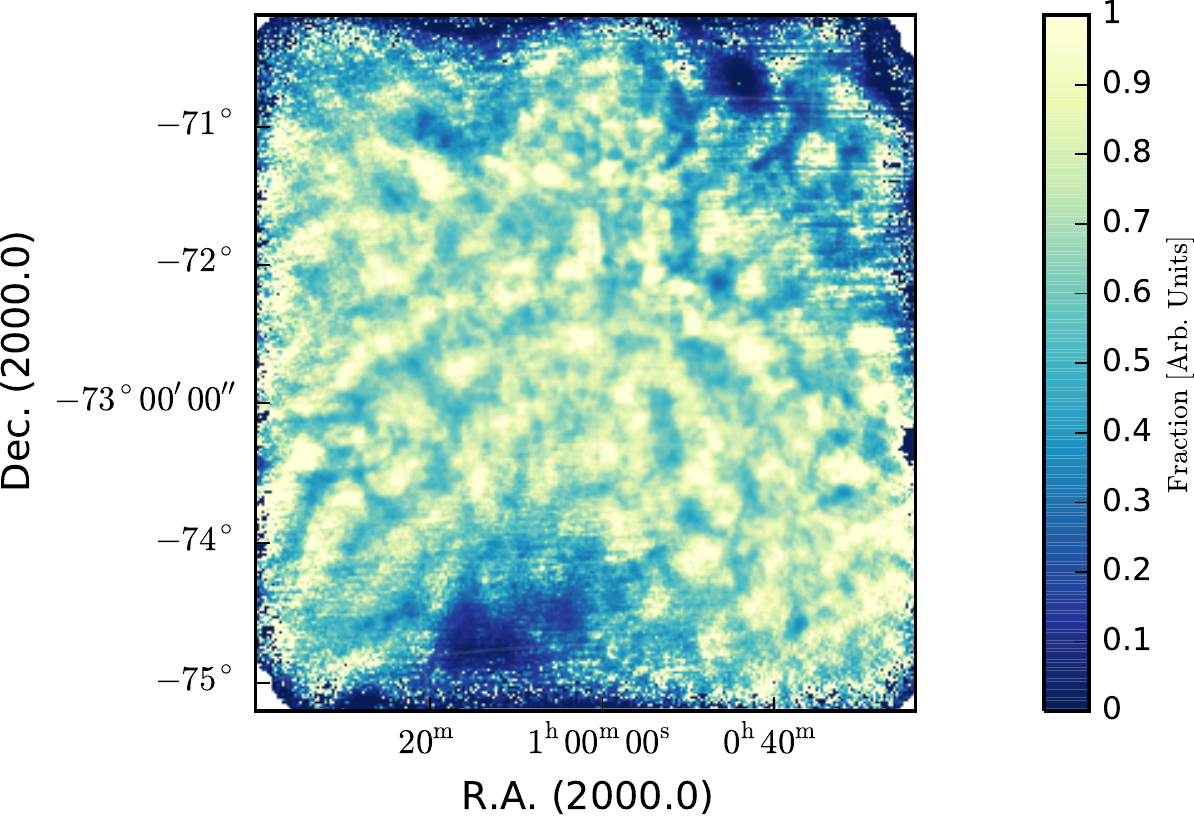}}
	\caption{The SMC flux density maps. Panel (a) shows the flux density map of the ATCA data, panel (b) the flux density map of the combined data. Panel (c) shows the the ratio of $\mathrm{missing\,spacings_{\,smc}} / \mathrm{combined_{\,smc}}$. It is the relative contribution of the emission gained through combination. It shows, how strongly the interferometric observation of the SMC suffers from the negative bowls.}
	\label{fig:smc_data}
\end{figure*}

Figure\,\ref{fig:smc_data} shows velocity integrated maps of the SMC. Panel (a) shows the flux density map of ATCA, panel (b) for the combination. Panel(c) quantifies the relative contribution of emission gained by the SSC. Apparently, there is a considerable amount of diffuse extended structures in the SMC, demonstrating how significantly the interferometric observation of the SMC suffers from the negative bowls \citep[][ their Fig.\,1a and b]{1985A&A...143..307B}. 

\begin{figure*}[hptb]
	\centering
	\subfigure[Cumulative Flux]{\includegraphics[width=0.95\columnwidth]{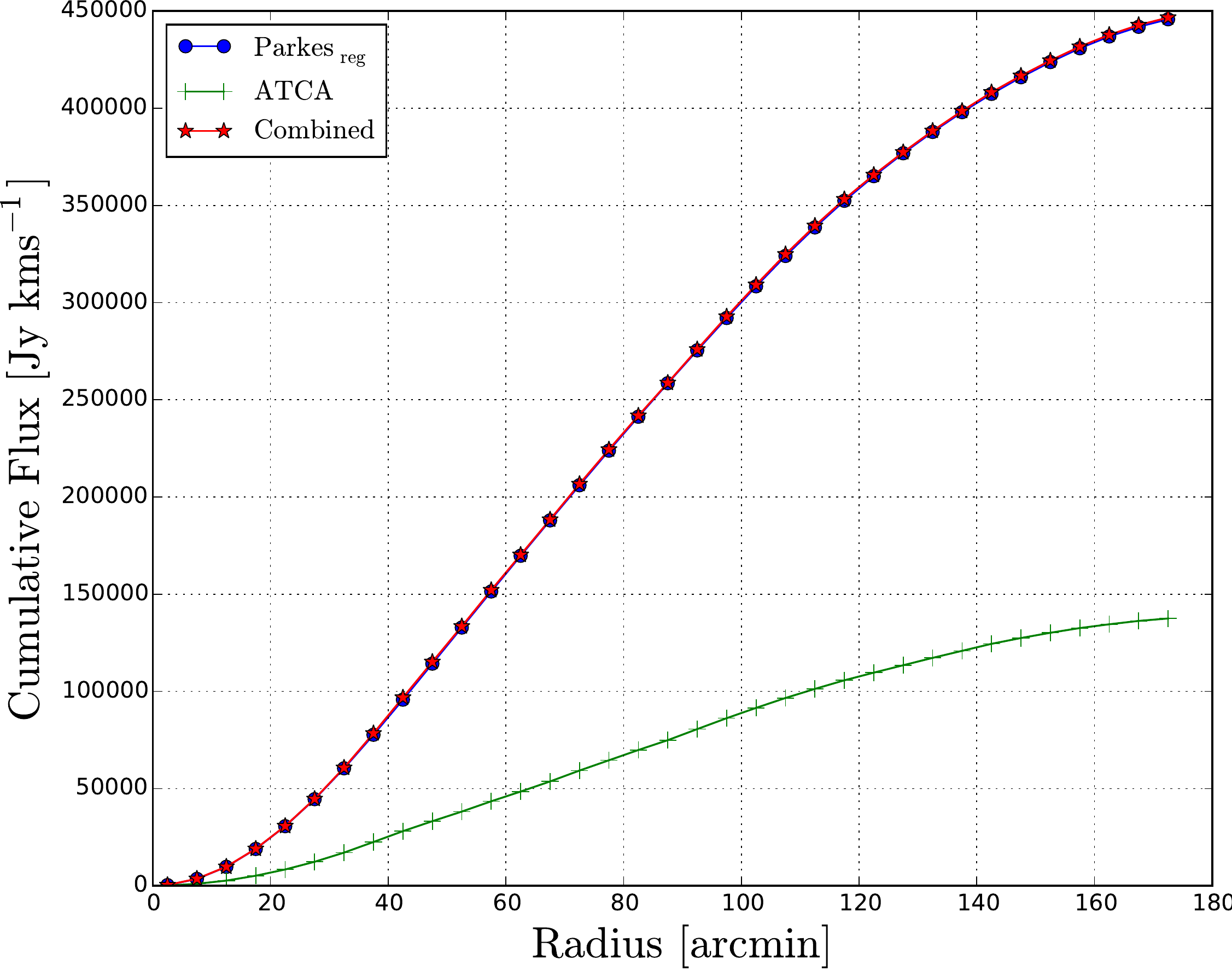}}
	\subfigure[Sum Spectra]{\includegraphics[width=0.95\columnwidth]{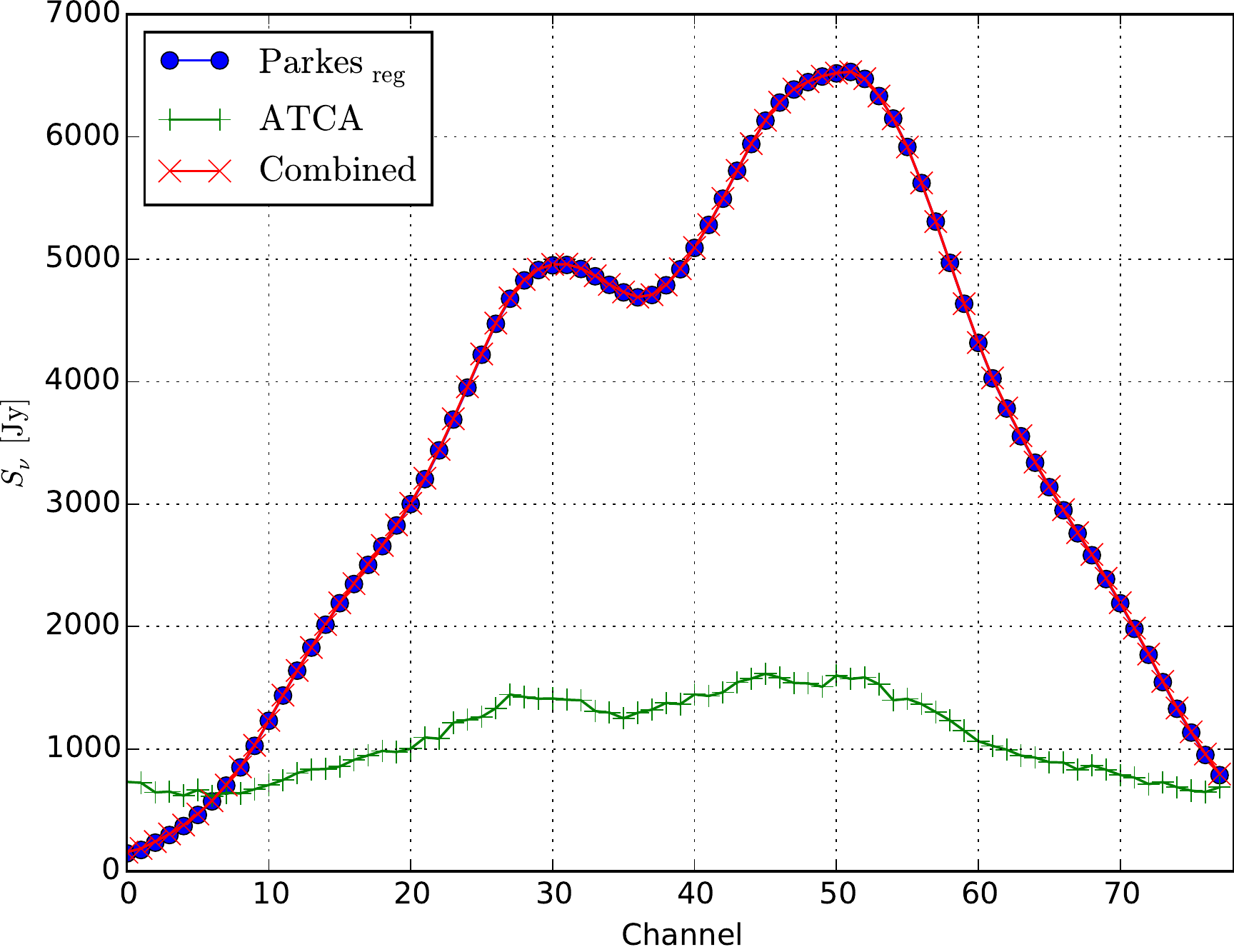}}
	\caption{Panel (a) shows the measured cumulative flux for the SMC data sets. Panel (b) presents the sum spectra of the same data sets, where the flux value is measured separately for each channel. In both panels the blue line presents the regridded Parkes data set, the green line the ATCA data, and the red line the combined data. Note that both red and blue points reveal very similar values as the measured values in single-dish and combined maps are in excellent agreement in both panels.}
	\label{fig:smc_annuli_spec}
\end{figure*}

Additionally, the cumulative flux as a function of radial separation from the center of the map (Fig.\,\ref{fig:smc_annuli_spec}, panel a) as well as sum spectra (Fig.\,\ref{fig:smc_annuli_spec}, panel b) for all three data sets are calculated. In both panels, the blue line presents the regridded low-resolution data ($\mathrm{Parkes_{\,reg}}$), the green line the high-resolution data (ATCA), and the red line the combined data. The measured cumulative flux density shows that the result of the combination is in line with the measured values from the regridded Parkes data for all radii. This is also true for the  flux density values in panel (b), where the total flux density is determined separately for each spectral channel. 
Note the strong deviation between the measured values in the first channels of the sum spectra (Fig.\,\ref{fig:smc_annuli_spec}, panel b) for single-dish and interferometric data sets. Here the interferometer receives significantly more flux than the single-dish. 
The origin of the deviation could not be conclusively determined from the data at hand. However, these first few channels are mainly noise dominated. The flux difference can be a result of using the MEM algorithm for cleaning the interferometer data cube as discussed in \citet{2002ASPC..278..375S}.

The combination results demonstrate the importance of the zero-spacing correction regarding determination of the physical and morphological properties of the objects. The results also show that the total flux in the combined map is quantitatively consistent with the total flux measured with the Parkes telescope, whereas the angular resolution of the ATCA data set is preserved.

\section{A Comparison of different SSC approaches}\label{sec:comp_method}

We present a comparison of the results of our combination method with two other common approaches. These are introduced in Sect.\,\ref{subsec:comb_sn} and \ref{subsec:comb_fea} respectively. The results for the SMC data set are discussed in Sect.\,\ref{subsec:comb_res}.

\subsection{Combination before deconvolution (CBD)}\label{subsec:comb_sn}

This method makes use of the linearity of the Fourier transform. The linearity allows to perform the SSC in the image domain. The result of the combination is a combined dirty image. The image needs to be deconvolved using an appropriate combined beam. The following equations describe the method mathematically:

\begin{eqnarray}\label{eq:ssc_method1}
I_\mathrm{\,comb}^\mathrm{D} & = & (I_\mathrm{\,int}^\mathrm{D} + \alpha \cdot f_\mathrm{\,cal} \cdot I_\mathrm{\,sd}^\mathrm{D}) / (1 + \alpha)\\
B_\mathrm{\,comb} & = & (B_\mathrm{\,int} + \alpha \cdot B_\mathrm{\,sd}) / (1 + \alpha)\nonumber.
\end{eqnarray}

$I_\mathrm{\,int}^\mathrm{D}$ is the interferometric dirty image, $I_\mathrm{\,comb}^\mathrm{D}$ the combined dirty image, and $B_\mathrm{\,comb}$ the combined synthesized beam. $\alpha$ estimates a factor for the resolution difference between the interferometric and single-dish data. $f_\mathrm{\,cal}$ is the measured calibration factor of the flux-density scales for the interferometric and single-dish data, where $f_\mathrm{\,cal}$ is retrieved from the overlap region of single-dish and interferometric data (compare Fig.\,1). It presents the systematic difference of calibration for interferometer and single dish. E.g., for the presented SMC data sets the value is $f_\mathrm{\,cal} = 1.05 \pm 0.05 $. 

For the combined data set, the deconvolution is performed in MIRIAD using the maximum entropy algorithm \citep[and references therein]{Stanimirovic_phd_1999}. 

This method requires both visibilities as well as the exact knowledge of single-dish and interferometric antenna pattern. The former is not a well determined quantity.

\subsection{Combination in the Fourier domain (Feather)}\label{subsec:comb_fea}

Feathering and its variations are the most commonly used approaches to perform the SSC, where the combination occurs in the Fourier (spatial frequency) domain. The \texttt{feather} task in CASA operates in a similar fashion as the \texttt{immerge} task in MIRIAD and \texttt{imerg} in AIPS \citep{2011ascl.soft07013I}. 

The combination method can be summarized as follows: First, both single-dish and imaged interferometric data cubes are Fourier transformed. Second, the Fourier transform of the regridded single-dish and interferometric data is tapered with two tapering functions $w'(k)$ and $w''(k)$. The sum of both tapering functions $w'(k)$ and $w''(k)$ is a Gaussian function with a FWHM value equal to that of the interferometric image \citep{2002ASPC..278..375S}. This ensures that the interferometric angular resolution is preserved after the combination. For the combination the single-dish data is deconvolved. The deconvolution is necessary since the single-dish data also has an antenna pattern. This can be retrieved by:

\begin{eqnarray}\label{eq:sd_vis}
	V(u,v) = \frac{V_\mathrm{\,sd}(u,v)}{b_\mathrm{\,sd}(u,v)},
\end{eqnarray}

where $V_\mathrm{\,sd}(u,v)$ is the \textit{single-dish visibilities}. In this case, the single dish is considered as an interferometer with infinite large number of receiving elements and a monotonically decreasing distribution of baselines from zero to $D_\mathrm{sd}$, where $D_\mathrm{sd}$ is the diameter of the single-dish telescope \citep{2002ASPC..278..375S}. $b_\mathrm{\,sd}(u,v)$ is the antenna pattern of the single dish. The resulting visibilities from Eq.\,\ref{eq:sd_vis} need to be rescaled by the scaling factor $f_\mathrm{\,cal}$ as described before. Hereafter the combination term is:

\begin{eqnarray}\label{eq:feathering_func}
V_\mathrm{\,comb}(k) = w'(k)\cdot V(k) + f_\mathrm{\,cal} \cdot w''(k) \cdot V_\mathrm{\,sd}(u,v). 
\end{eqnarray}

After combination the result is transformed back to the image domain \citep{Stanimirovic_phd_1999}. For feathering it is important that the input images have a well-defined beam shape. The deconvolution step (Eq.\,\ref{eq:sd_vis}) also requires care. The Fourier transform of the beam approaches zero for large values of $k$. As a result of this, the noisier high spatial frequencies are even amplified. To minimize this effect, an appropriate tapering function is to be applied. 

We used the \texttt{feather} task in CASA to perform the combination in the Fourier domain. The inputs are the regridded single-dish and interferometric data, observed with 64 m Parkes telescope and the ATCA (Sect.\,\ref{sec:smc_data}). The applied scaling factor $f_\mathrm{\,cal}$ for the single-dish data is $\sim 1$ (Sect.\,\ref{subsec:comb_sn}).  

\subsection{Results of combination for the SMC data sets}\label{subsec:comb_res}

\begin{figure*}[phtb]
	\centering
	\includegraphics[width=0.98\textwidth]{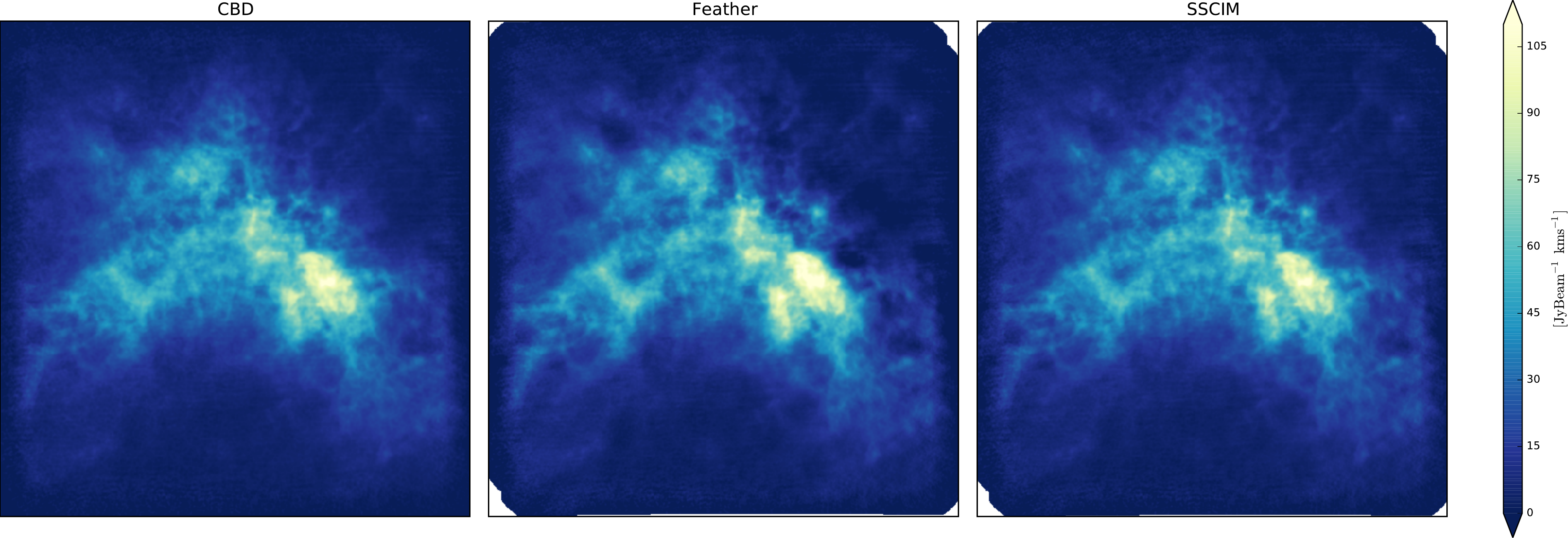}
	\caption{Velocity-integrated maps for all three combination methods. The left panel is the result of combination before deconvolution (CBD), the middle map the combination in the spatial frequency domain (feather), and the right panel our combination method (SSCIM).}
	\label{fig:smc_comb_mom0}
\end{figure*}

Figure\,\ref{fig:smc_comb_mom0} shows the velocity-integrated maps for all three combination methods. The left panel is the CBD result, the middle map the combination in the spatial frequency domain applying feathering, and the right panel the result of our combination approach.

The amount of recovered integrated flux density from the velocity-integrated maps is very similar in all three cases. The values are consistent with the corresponding value from the regridded single-dish data. Both, CBD and feathering are sensitive to bright isolated structures within the primary beam and close to its rim. Thus, the single-dish data needs to be tapered prior to combination \citep{Stanimirovic_phd_1999}. This leads to a lower final angular resolution of the scientific data. This is not mandatory for our combination method, since this method does not require any Fourier transform.

\begin{figure}[!tb]
	\centering
	\includegraphics[width=0.95\columnwidth]{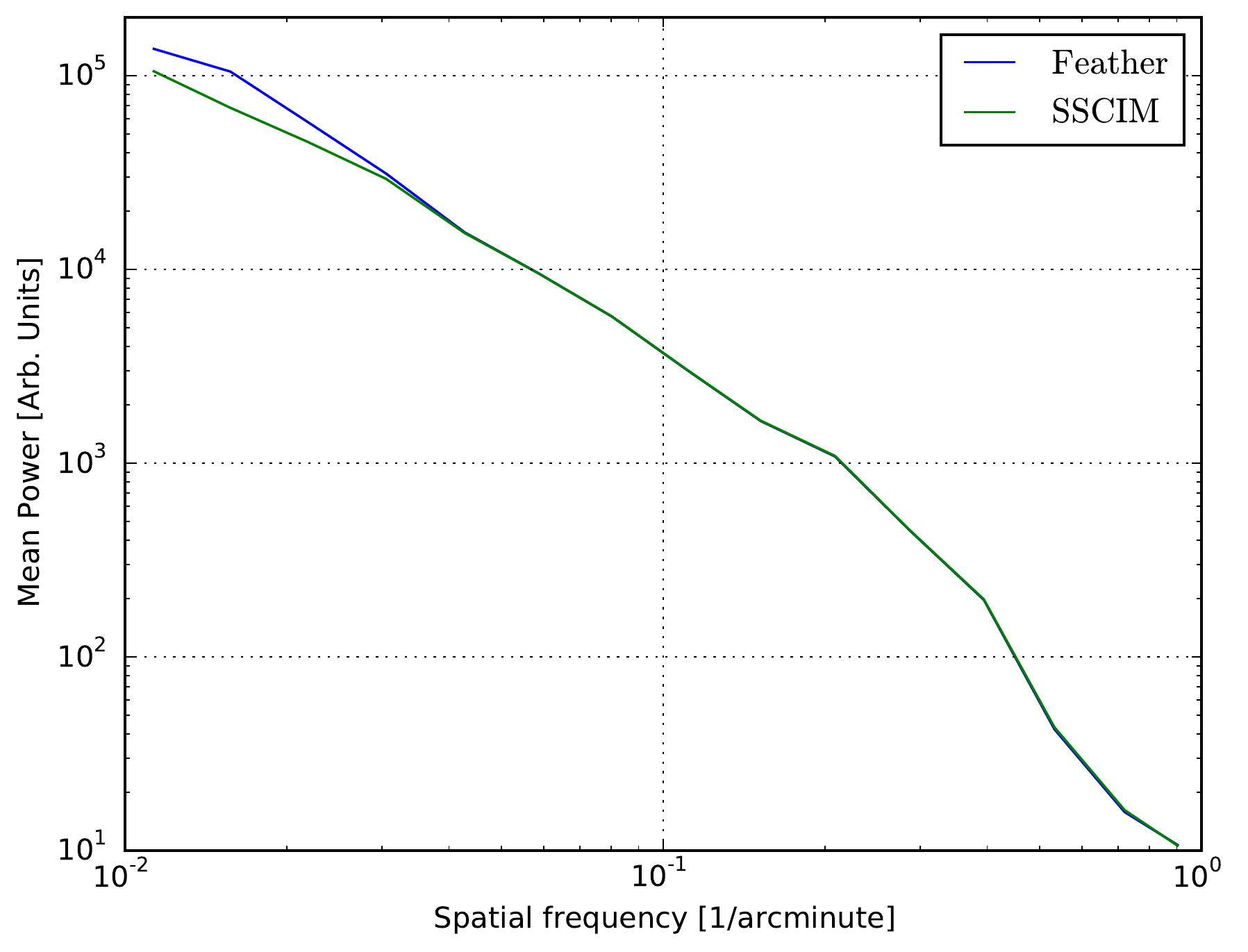}
	\caption{Power spectral density (PSD) profiles of two combined data sets. The blue line represents the PSD for the combination using feathering, the green line for SSCIM.}
	\label{fig:smc_psd_prof}	
\end{figure}

Figure\,\ref{fig:smc_psd_prof} shows the power spectral density (PSD) profiles of two combined data sets. The green line shows the PSD profile of the combined data set using our introduced approach, the blue line for the combined data set using Feather task in CASA, respectively. Both methods show very similar results at middle and higher spatial frequencies. However, there exists a difference in the amount of measured power at the lower spatial frequencies, where feathering shows higher values at these regions. It is unclear what causes this difference. But since in feathering tapering and deconvolution is involved, we think that these operations may change the flux at the largest scales. It might be related to both w' and w'' parameters as introduced in Eq.\,\ref{eq:feathering_func}. 

\section{Synthesized imaging parameters} \label{sec:syn_image_par}

An interferometric image is a \textit{model} or \textit{best guess} of the true brightness distribution of the object. Different choices and strategies during the process of imaging affect the final result substantially. In the current section we discuss the impact of two imaging parameters: pixel size and weighting scheme on the reduced interferometric data.

\subsection{Pixel size} 

For the Fourier transformation the visibilities need to be brought onto a regular grid. This is realized by convolving the raw data with a specific gridding kernel.
We chose to Nyquist sample the data and thus the pixel size is set to $\approx 1/(2\cdot\sqrt[]{2})$ of the FWHM \citep{winkel_phd_2008}.

\subsection{Weighting scheme} 

For an interferometric observation, the density of the sampling points in the $(u,v)$-plane is not uniform and varies with the observing time. The coverage of the central regions of the $(u,v)$-plane is commonly more complete because of redundancy than in its outer regions. Different weighting schemes have been introduced to emphasize different regions in the $(u,v)$-plane \citep{briggs_phd}. For a specific weighting scheme, Eq.\,\ref{eq:dirty_im} can be modified as follows:

\begin{eqnarray}\label{eq:weighted_dirty_im}
\mathfrak{F}^{-1}(V^{(v)}_\mathrm{obs}(u,v)) = \\
\mathfrak{F}^{-1}(V^{(v)}_\mathrm{true}(u,v) \cdot S^{(v)}(u,v) \cdot W(u,v)) \nonumber
\end{eqnarray}

where $W(u,v)$ describes the applied weighting scheme. Consequently, the dirty beam and clean beam change through multiplication with $W(u,v)$. 

For interferometric observations, each visibility sample is given a weight during the imaging process (see Eq.\,\ref{eq:weighted_dirty_im}).Different weighting schemes give a trade off between higher sensitivity and higher angular resolution \citep{briggs_phd}. In the following we briefly introduce some weighting schemes:

\textit{Natural} weighting emphasizes all visibilities equally. For this weighting scheme $W(u,v) \propto 1$. 

In the \textit{Uniform} weighting, weights are inversely proportional to the density $N$ of the sampling function $W(u,v) \propto 1/N$. The latter weighting scheme emphasizes the less sampled long baselines resulting in a higher noise level in the reconstructed interferometric image. This latter scheme, however, yields a better resolution, i.e., smaller synthesized beam, compared to that of natural weighting. 

\textit{Robust} weighting parameterizes the weighting function with a single parameter $R$ to vary between the natural and uniform weighting schemes. By varying this parameter, images with sensitivities close to naturally weighted maps but with angular resolutions closer to those of uniform weighting \citep{briggs_phd} are calculated.

\section{Impact of imaging parameters on the interferometric and SSC data}\label{sec:impac_imag_par}

In this section, we discuss the influence of the visibility weighting scheme and pixel size on the resulting interferometric image and SSC. 

\begin{figure*}[hpbt]
	\centering
    \includegraphics[width=0.99\textwidth]{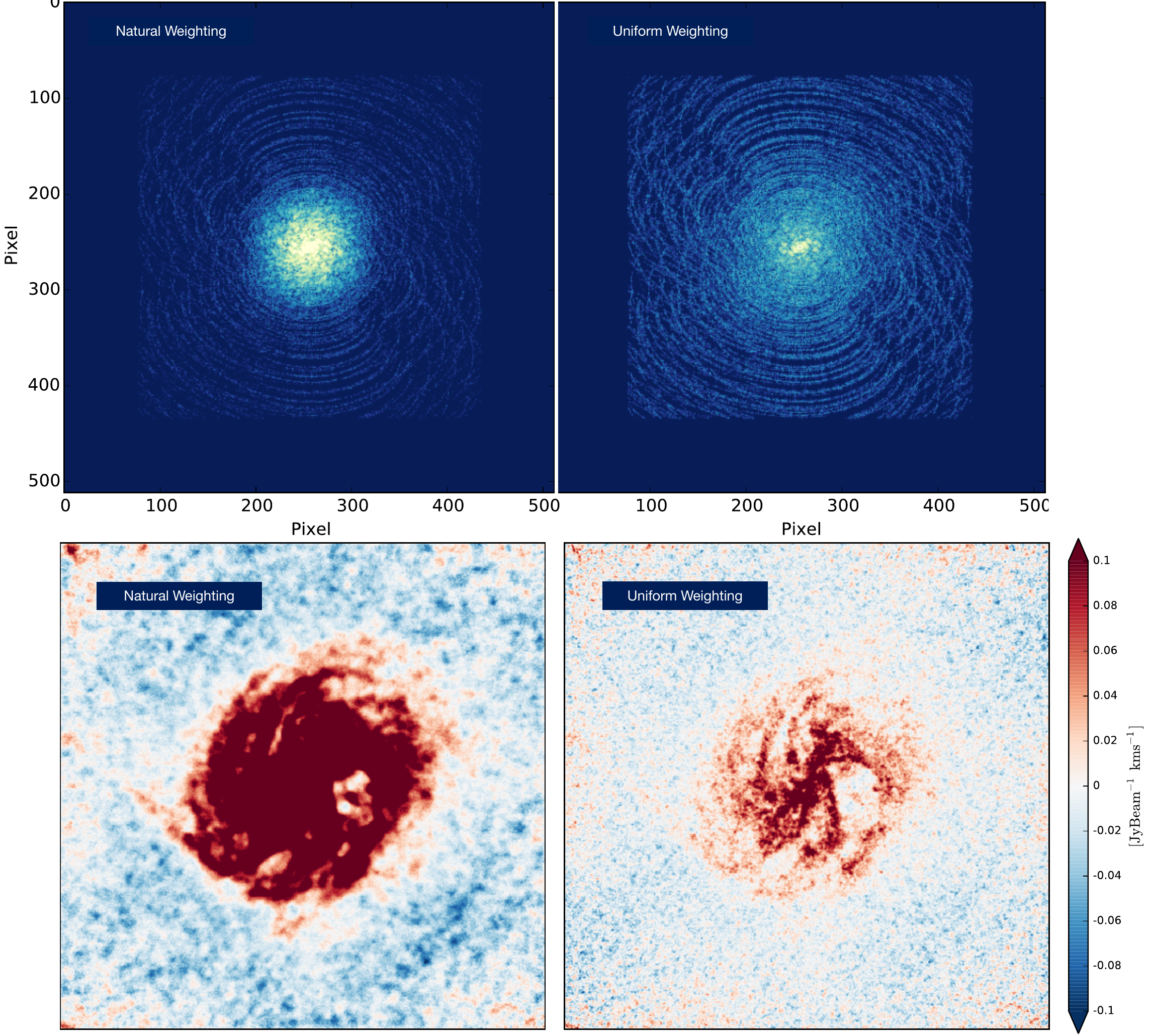}
	\caption{Effect of different weighting schemes on the resulting sampling function and interferometric image. The top panels show the result of a Fast Fourier Transformation (FFT) of the cleaned interferometric observations of NGC\,4214 with different weighting schemes. The top left panel shows the gridded $(u,v)$-coverage for natural weighting, the top right panel for uniform weighting. The bottom panels show the velocity-integrated maps of the same observation for the natural and uniform weighting schemes, respectively. The chosen pixel size for this observation is 3 arcseconds. For these observations, the measured beam size for the naturally weighted data set is about 20 arcseconds. The corresponding value for the uniformly weighted data are about 10 arcseconds.}
	\label{fig:weighting_scheme}	
\end{figure*}

Figure\,\ref{fig:weighting_scheme} demonstrates the effect of different weighting schemes on the resulting sampling function as well as the final interferometric image for the NGC\,4214 data sets. The presented results stem from the same visibilities and have the same FoV, however, they differ in the applied weighting schemes. The arrangement of the figure is as follows: The top panels show the result of a Fast Fourier Transformation (FFT) of the cleaned interferometric observations of NGC\,4214 with different weighting schemes. The top left panel shows the gridded $(u,v)$-coverage for natural weighting, the top right panel for robust weighting.The bottom panels show the velocity-integrated maps of the same observation for the natural and uniform weighting schemes, respectively. The chosen pixel size for this observation is 3 arcseconds. In the top left panel the more numerous visibilities at small $(u,v)$-distance lead to a lower rms noise level and higher sensitivity towards large angular scale structures, whereas the applied weighting scheme in the top right panel increases the weight for visibilities at large $(u,v)$-distance resulting in a higher noise level. The latter weighting scheme puts the emphasis on the small angular scale structures and achieves a better angular resolution compared to the former one. The measured beam size for the naturally weighted data set is about 20 arcseconds, whereas the corresponding values for the uniformly weighted data is about 10 arcseconds.

\subsection{Flux variations as a function of weighting scheme and pixel size}\label{subsec:flux_var}

\begin{figure}[hptb]
\centering 
\subfigure[NGC\,4214: cumulative flux]{\includegraphics[width=0.95\columnwidth]{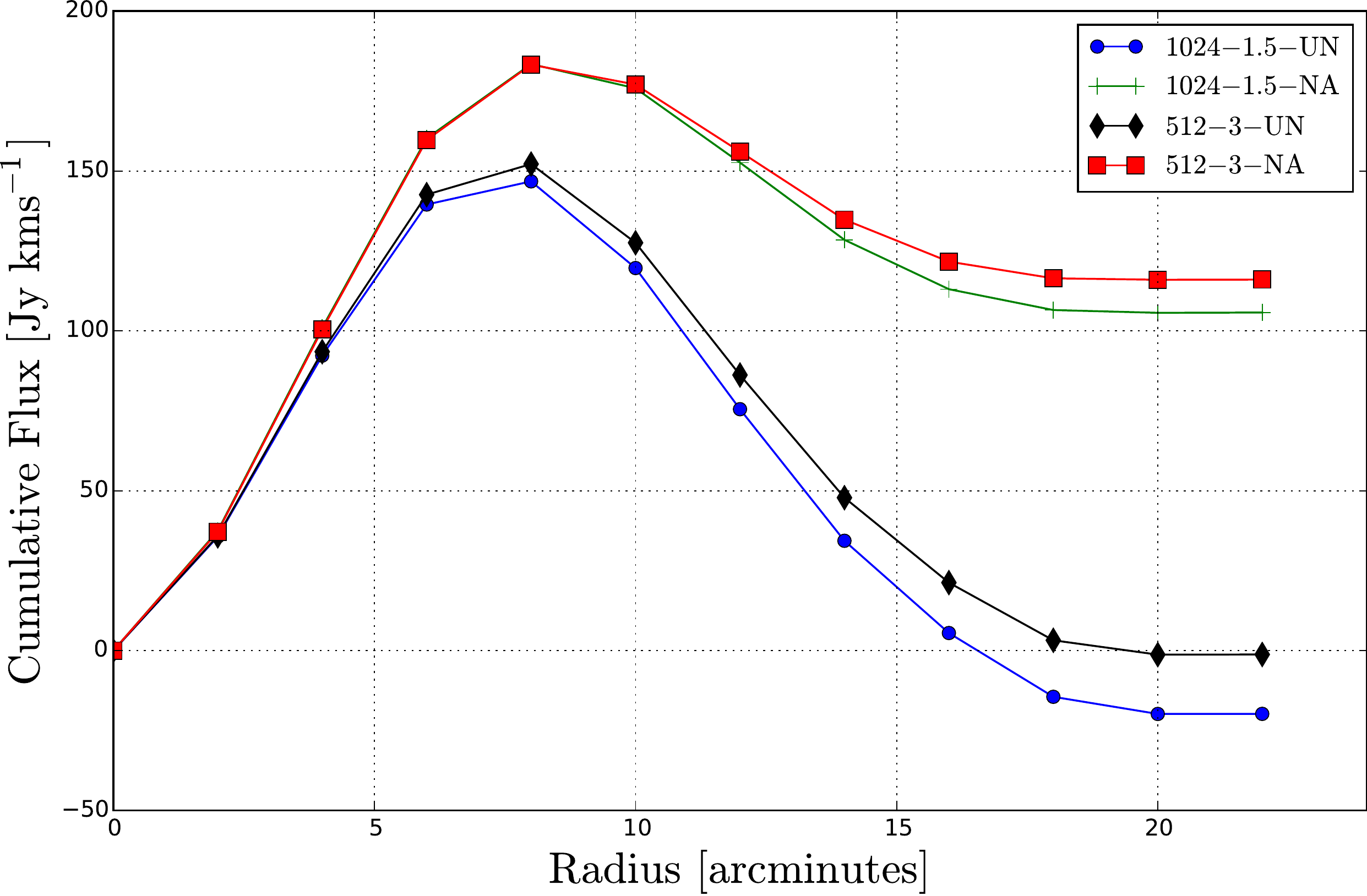}}
\subfigure[NGC\,4214: PSD profiles]{\includegraphics[width=0.95\columnwidth]{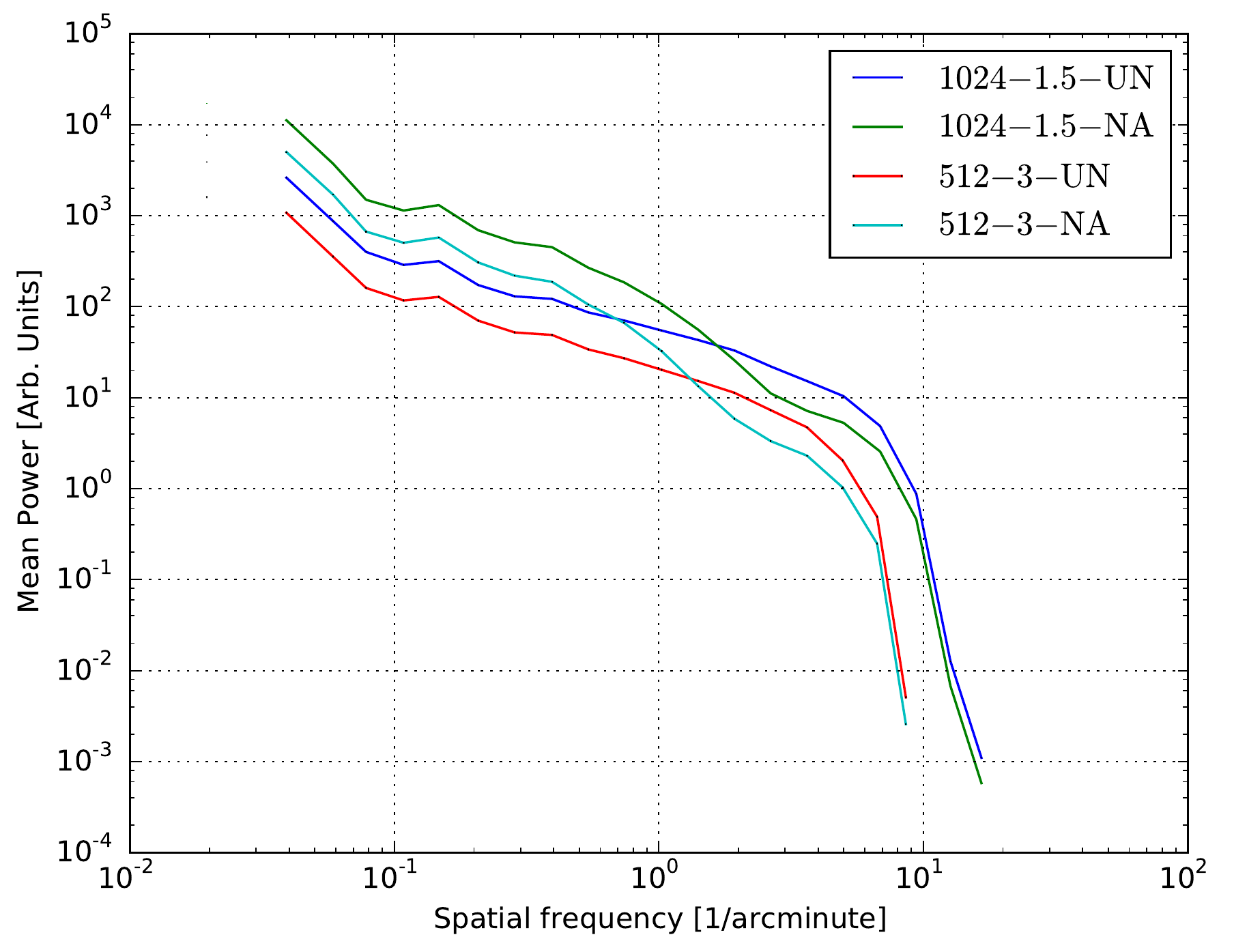}}
\caption{Measured cumulative flux as a function of radial separation from the center of the velocity-integrated maps and PSD profiles for NGC\,4214. NA and UN stand for natural (robust 5) and uniform (robust 0.5) weighting, respectively. Two different pixels sizes of 1.5'' and 3'' have been considered. 512 and 1024 are the number of pixels along both spatial axes. Panel (a) shows the flux profiles of all the cleaned interferometric data sets of NGC\,4214. Panel (b) shows the PSD profiles of all NGC\,4214 data sets. The blue and red lines show the result of PSD for the uniformly weighted data sets with a pixel size of 1.5'' and 3'', respectively. The green and cyan lines for the naturally weighted data with a pixel size of 1.5'' and 3'', respectively.}
\label{fig:n4214_annflux_psd}
\end{figure}

\begin{figure}[!tb]
	\centering
	\includegraphics[width=0.95\columnwidth]{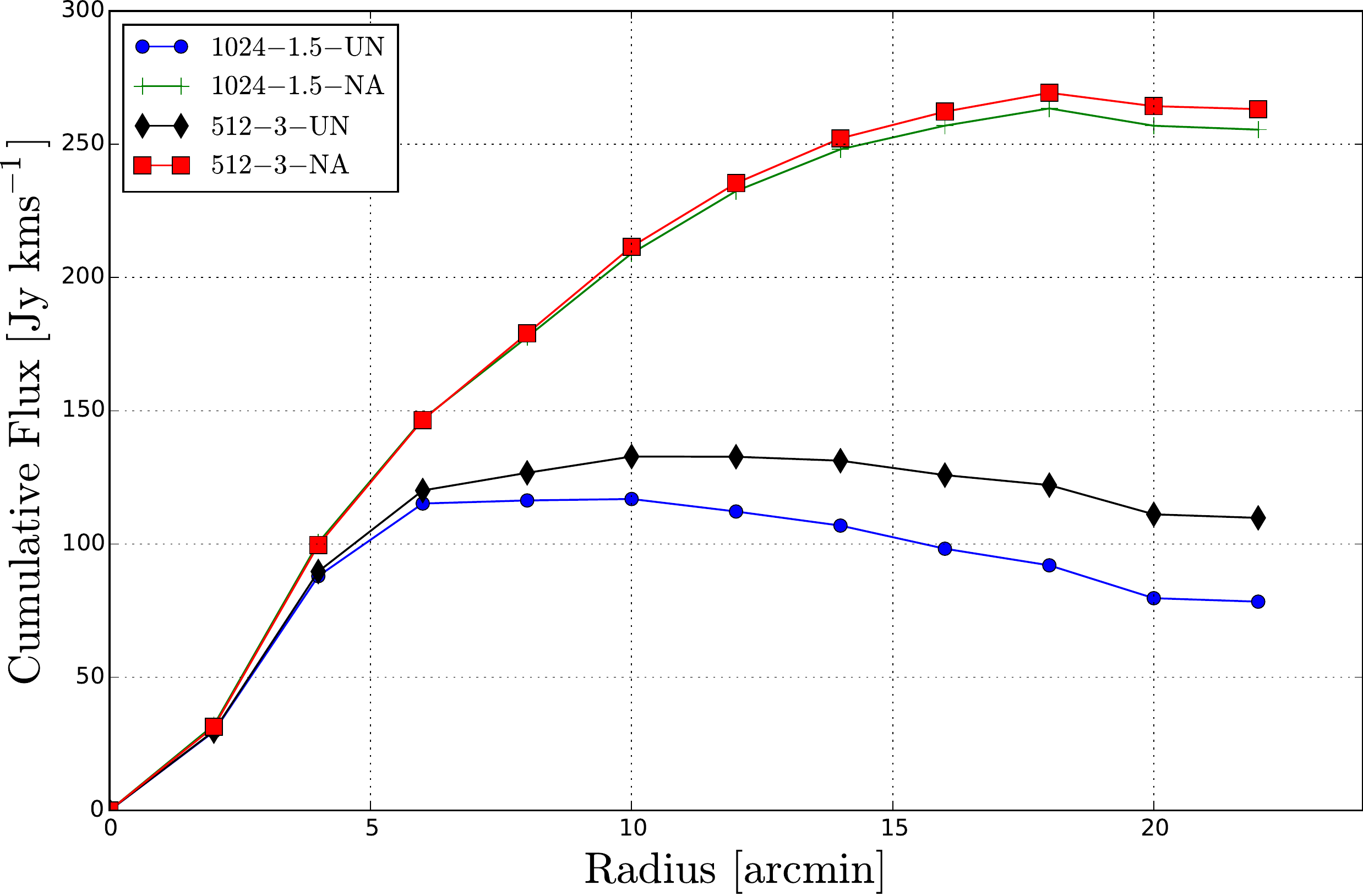}
	\caption{Measured cumulative flux as a function of radial separation from the center of the unmasked velocity-integrated intensity maps of NGC\,5055. NA and RO stand for natural and robust weighting, respectively. Two different pixels sizes of 1.5'' and 3'' have been considered. 512 and 1024 are the number of pixels along both spatial axes. Note that the impact of applied weighting scheme (sensitivity) is more significant compared to the impact of chosen pixel size for these flux profiles. However, the course of the curves for the NGC\,5055 data sets are quite different compared to those of NGC\,4214 data sets (panel (a) of Fig.\,\ref{fig:n4214_annflux_psd})}
	\label{fig:n5055_annflux}	
\end{figure}

For all interferometric data sets of NGC\,4214 and NGC\,5055, the cumulative flux as a function of radial separation from the center of the map in ever larger radii is measured. Thus, the measured value in the largest radius corresponds to the integrated flux density in each data set. Panel\,(a) of Fig.\,\ref{fig:n4214_annflux_psd} shows the measured flux densities for all data sets of NGC\,4214, Fig.\,\ref{fig:n5055_annflux} for NGC\,5055, respectively.  

For the NGC\,4214 data sets the measured total flux increases in the central regions. The maximum value is measured at a radius of about 8 arcminutes for all 4 data sets. At this radius the bulk of the emission from the galaxy is measured. For the larger radii the cumulative flux decreases. This is due to the deep negative bowls around the galaxy, which are the result of the missing spacings as described in Sect.\,\ref{subsec:ssp}. Note the significant difference in the measured total fluxes for both data sets with the natural weighting (panel a - red and green lines) compared to the values measured in data cubes with the robust weighting (panel a - blue and black lines). For a given weighting scheme, the amount of measured total flux is higher for data sets with larger pixel size (in this case, 3 arcseconds).

The effect of applied weighting scheme can be summarized as follows: The uniform (robust 0.5) weighting scheme puts the emphasis on the long baselines. The sampling points in these regions of the $(u,v)$-coverage are more sparse than the central regions. This results in a higher noise level and lower sensitivity towards large-scale structures. Additionally, the amplitude of the sidelobes in the synthesized antenna pattern are higher which results in deeper negative bowls around the structure. Note, the difference in the measured total flux for data sets with natural and uniform weighting, where the measured total fluxes in the naturally weighted data sets are significantly higher than those of the uniformly weighted data sets.

The overlap region between single-dish and interferometric observation decreases significantly for uniform weighting. This is an important factor for the combination. Therefore, are uniform weighted data sets are less appropriate for the combination.

Note, that the effect of different pixel sizes compared to the choice of applied weighting scheme is smaller but not negligible. The analysis demonstrates that the effect of pixel size is purely a smoothing effect. A larger pixel grid smoothes the visibilities. Panel (b) of Fig.\,\ref{fig:n4214_annflux_psd} shows the result of PSD for different NGC\,4214 data sets. The PSD profiles show that we measure higher power at lower spatial frequencies for naturally weighted data set for a given pixel size, corresponding to higher power at larger angular scales. The trend, however, changes at higher spatial frequencies, where the uniformly weighted data sets (with their smaller synthesized beam) recover more emission. The smoothing as a result of a larger pixel size suppresses both low and high spatial frequencies, where the PSD profiles reveal less power (red and cyan lines compared to blue and green lines). The result also shows that smoothing affects the measured total flux, however, it is scale independent. 
 
The smoothing decreases the amplitude of the negative bowls around the bright structures (Sect.\,\ref{subsec:ssp}). Therefore, the measured total fluxes for data sets with larger pixel size are higher for a given weighting scheme (panel (a) of Fig.\,\ref{fig:n4214_annflux_psd}). Overall, our case study provides an idea of the magnitude of the effect of changing pixel size or weighting scheme on the final flux distribution in these data sets.

Figure\,\ref{fig:n5055_annflux} shows the result of flux profiles for NGC\,5055. The curves are quite different for the NGC\,5055 data sets compared to those of NGC\,4214 data sets (panel (a) of Fig.\,\ref{fig:n4214_annflux_psd}). The most significant difference is the strong increase of the measured total flux values for naturally (robust 5) weighted data sets, whereas the corresponding values are constant for the uniformly (robust 0.5) weighted data. This result suggest that the interferometric array yields more information regarding different scales and therefore, the negative bowls are flatter compared to those of NGC\,4214.

It is important to mention that the position as well as the shape and depth of the negative bowls depends on distance, extent, and orientation of the galaxy on the sky as well as on the antenna pattern of the interferometric array. The negative bowls arise when the interferometer lacks large-scale information. NGC\,5055 is located at a larger distance compared to NGC\,4214 (Table\,\ref{tb:n4214_n5055_obs_par}). This suppresses the amplitude of the negative bowls. This is the reason why the drops obvious in Fig.\,\ref{fig:n5055_annflux} are smaller than those present in panel (a) of Fig.\,\ref{fig:n4214_annflux_psd}.

\subsection{SSC for NGC\,4214}

The current section presents the result of the combination for all four NGC\,4214 data sets as described in Sect.\,\ref{subsec:flux_var}. The SSC for NGC\,5055 will be presented in a follow-up paper, including yet deeper and more extended \hi observations from the HALOGAS survey \citep{2011A&A...526A.118H}.

The SSC is performed using the combination method as described in Sect.\,\ref{sec:pipeline}. The missing spacings are provided by the Effelsberg Bonn \hi Survey \citep[EBHIS, ][]{2010ApJS..188..488W, 2011AN....332..637K, 2016A&A...585A..41W}. The angular resolution of the EBHIS data cube is 10.8', the corresponding values in the VLA and combined data sets vary approximately between 6'' and 20''. The spectral resolution of the regridded EBHIS, VLA and combined cubes is about 1.3 \kms.

\begin{figure*}[hbtp]
	\centering
	\subfigure[NGC\,4214 - VLA]{\includegraphics[width=0.43\textwidth]{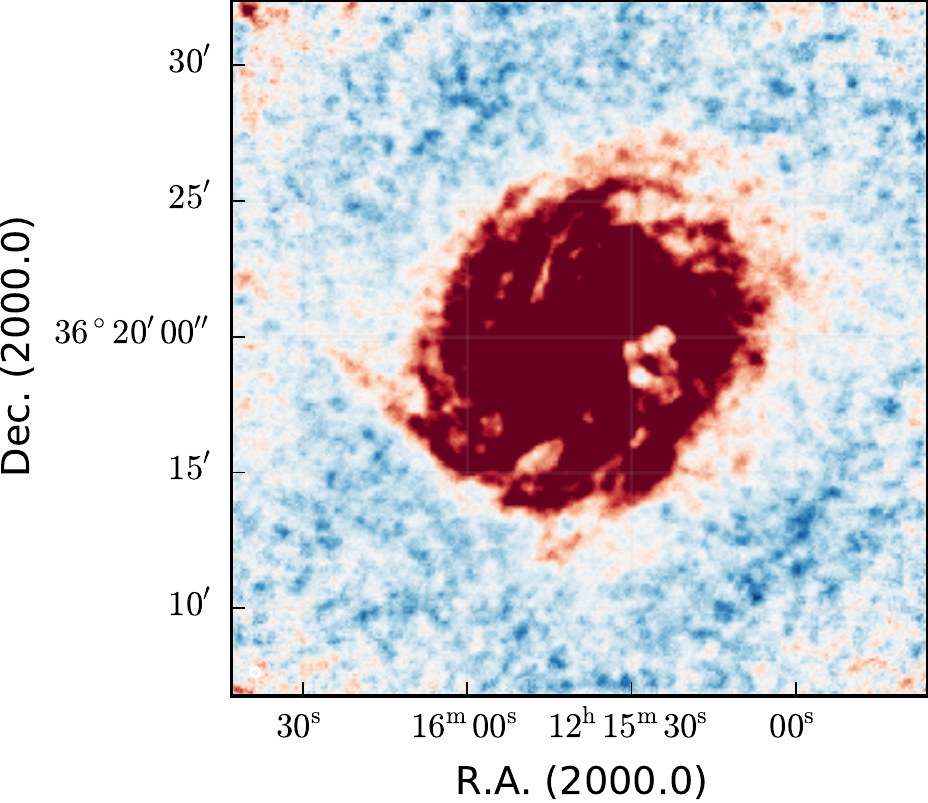}}
	\subfigure[NGC\,4214 - Combined]{\includegraphics[width=0.56\textwidth]{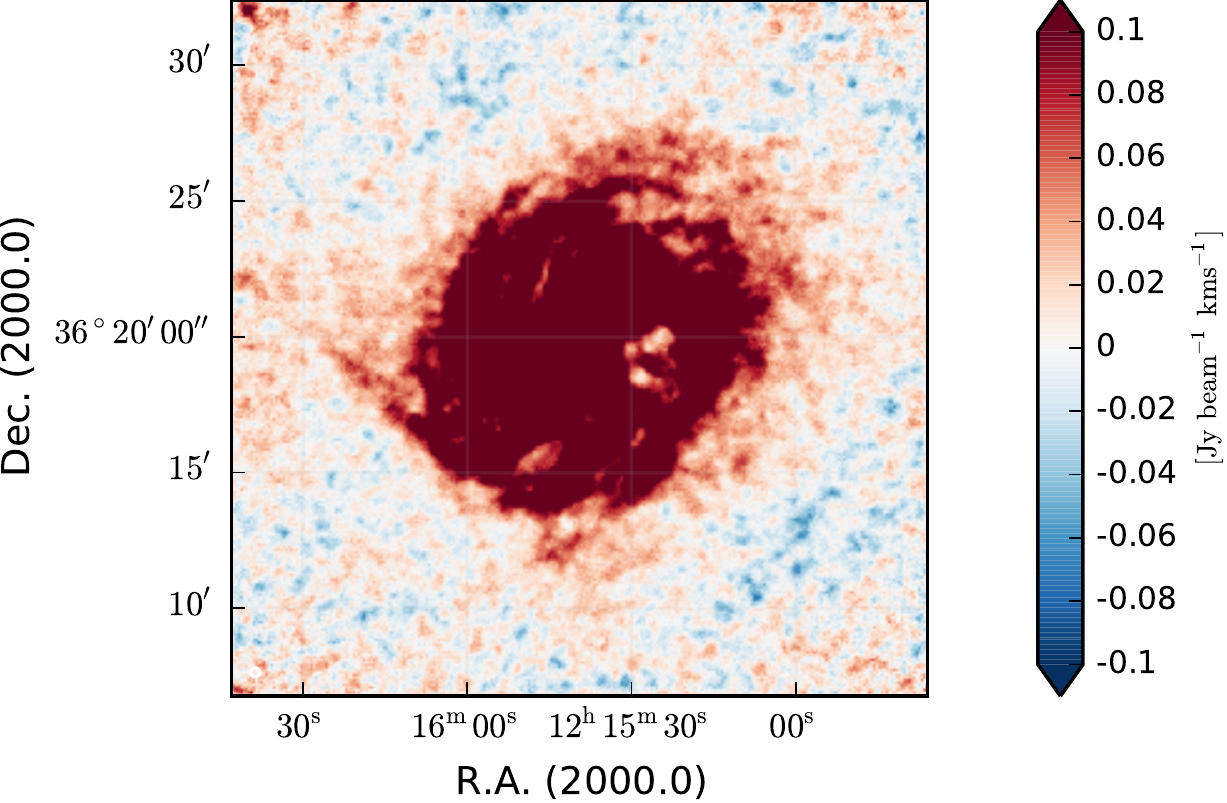}}
	\caption{Velocity-integrated maps of NGC\,4214. Panel (a) shows the VLA observations of the galaxy, panel (b) the result of the combination with the EBHIS single-dish data. The interferometric data are naturally weighted with a pixel size of 3''. Note the strong negative bowls around the galaxy obvious in panel a (blue colors).}
	\label{fig:n4214_ssc}
\end{figure*}

Figure\,\ref{fig:n4214_ssc} shows the result of combination for the natural weighted NGC\,4214 data set with a pixel size of 3''. The left panel shows the velocity-integrated VLA map, the right panel that of combination, respectively. Note the significance of the negative bowls around the structure obvious in panel (a). 

\begin{figure*}[htpb]
\centering
\subfigure[NGC\,4214 - 512 3'' NA]{\includegraphics[width=0.45\textwidth]{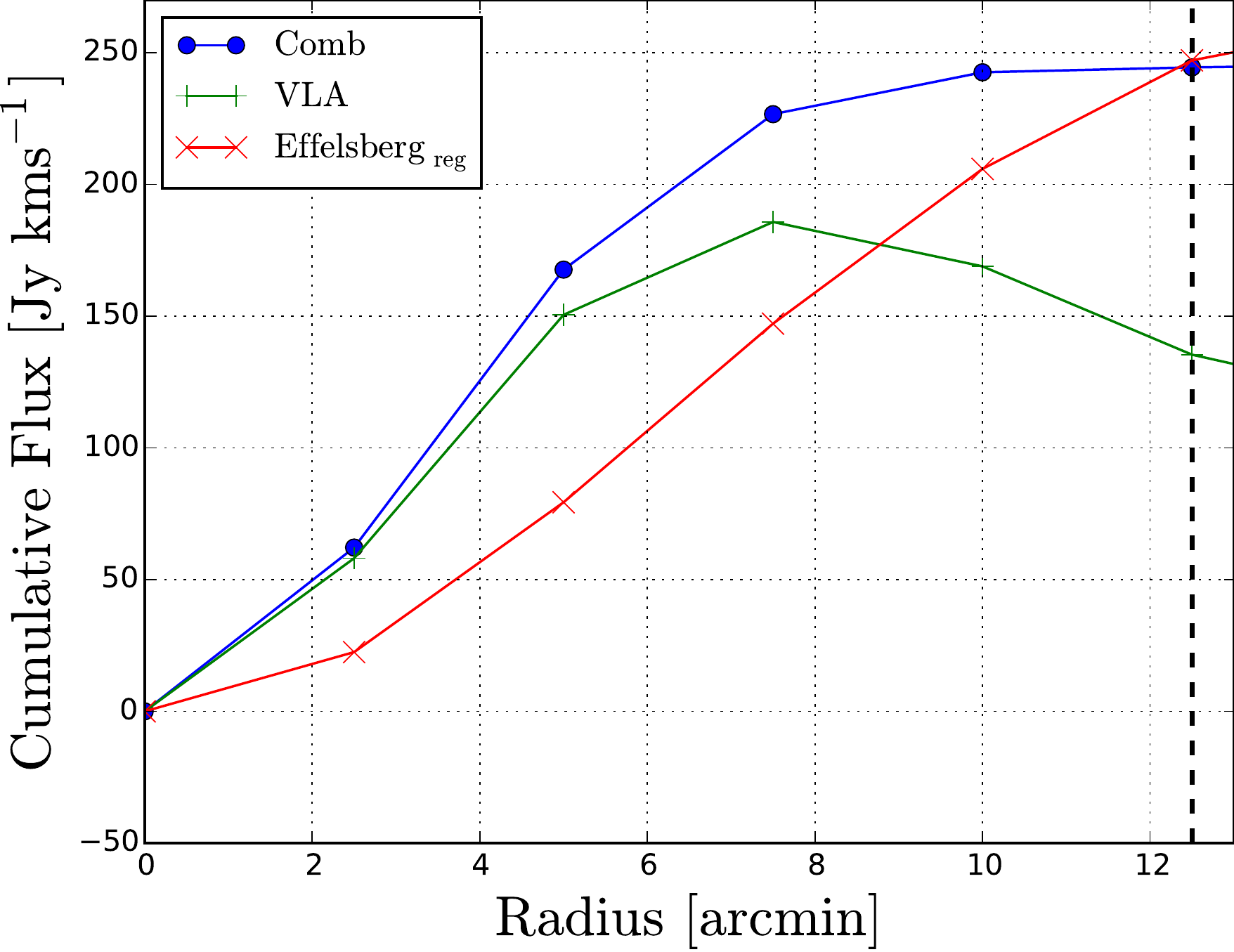}}
\subfigure[NGC\,4214 - 512 3'' UN]{\includegraphics[width=0.45\textwidth]{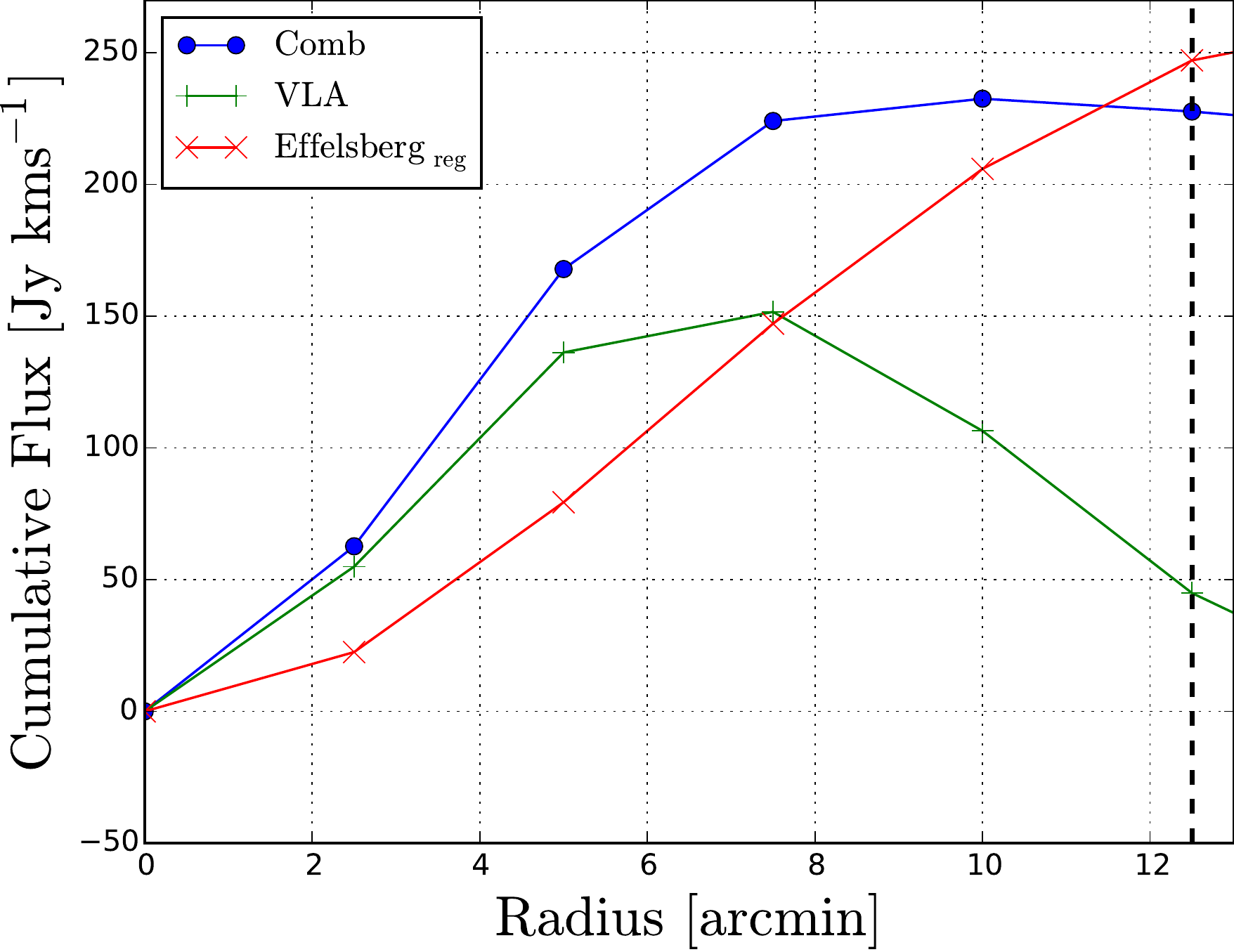}}
\subfigure[NGC\,4214 - 1024 1.5'' NA]{\includegraphics[width=0.45\textwidth]{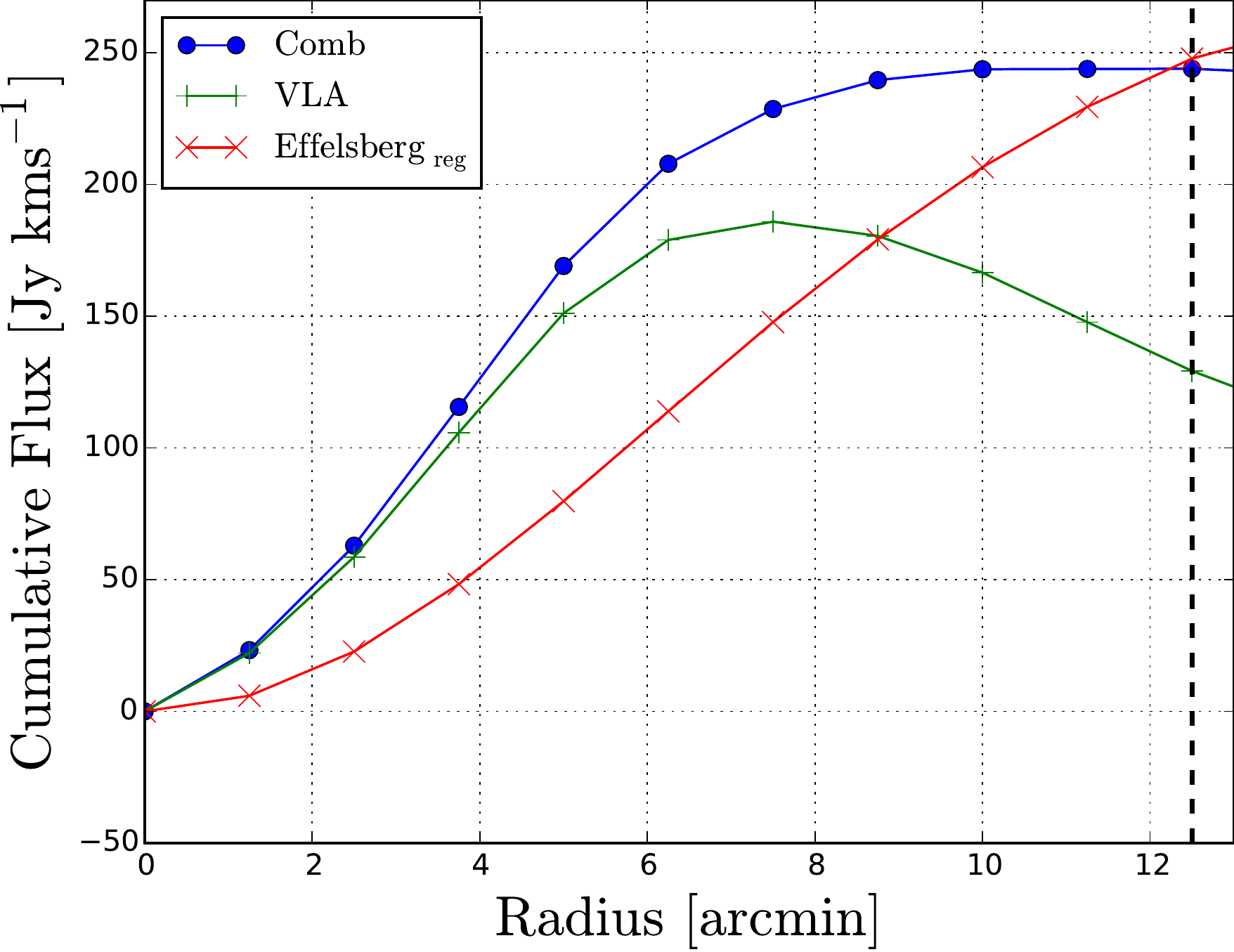}}
\subfigure[NGC\,4214 - 1024 1.5'' UN]{\includegraphics[width=0.45\textwidth]{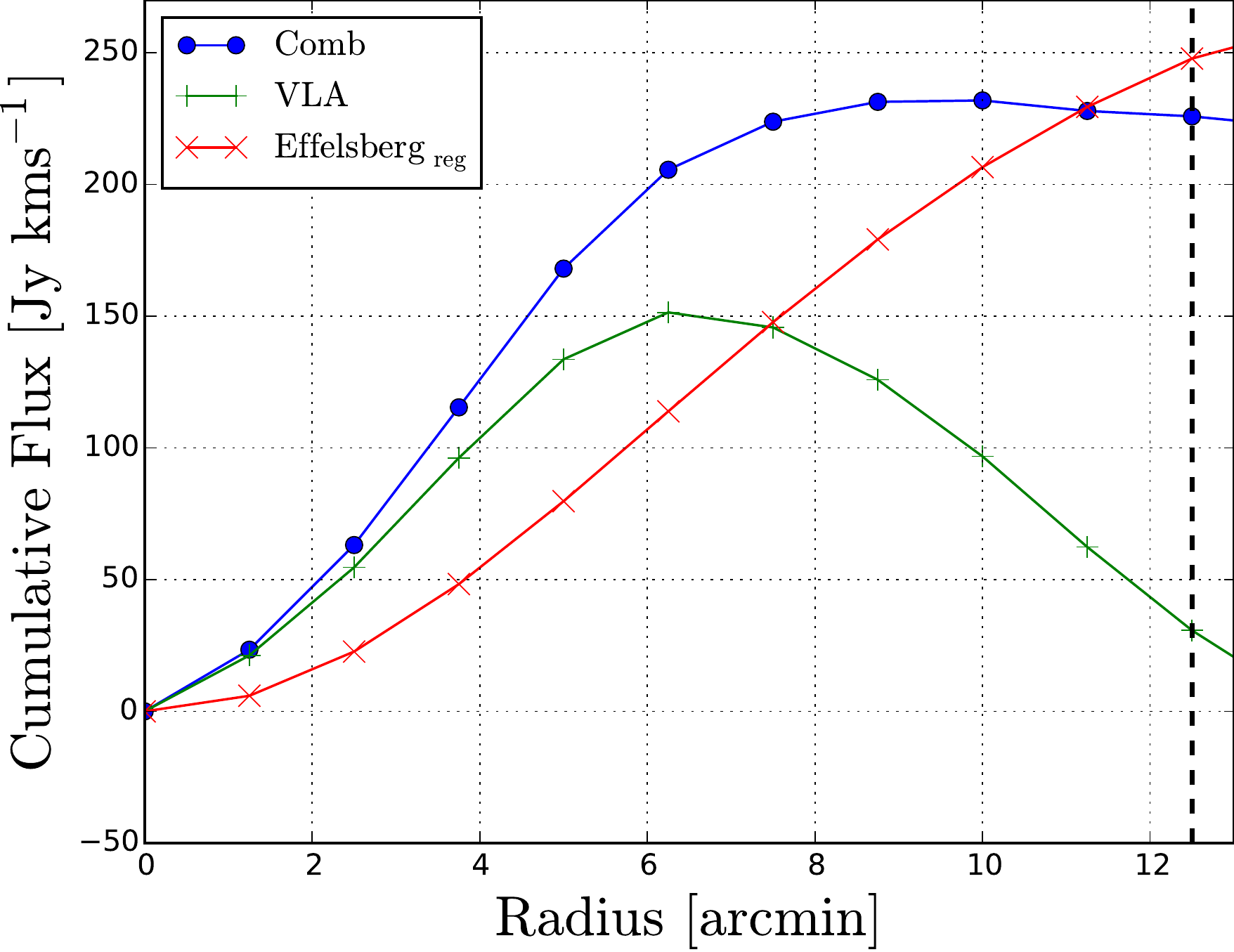}}	
\caption{Measured cumulative fluxes as a function of radial separation from the center of the map. In all the panels the blue represents the measured fluxes for the combination, the green line the values for the VLA data, and the red line the values for the regridded EBHIS data. The dashed line at 12.4' corresponds to the largest probed radius in the data sets. The number of pixels along the spatial axes varies between 512 and 1024. The chosen pixel sizes are 1.5'' and 3''. NA stands for natural weighting, UN for uniform.}
\label{fig:n4214_ssc_annuli}
\end{figure*}

Figure\,\ref{fig:n4214_ssc_annuli} shows the measured cumulative fluxes for the combined (blue), VLA (green), and regridded Effelsberg (red) data cubes of the different NGC\,4214 data sets. The maximum radius probed by the observation is 12.4' and is marked with a dashed line. We distinguish between naturally and uniformly weighted data with designations NA and UN, respectively.

For both naturally weighted data sets, the amount of measured flux densities at the largest radius, i.e., the accumulated flux across the entire map is in good agreement with the corresponding value measured in the regridded EBHIS data. For the uniformly weighted data, the measured flux densities in the combined map are smaller than the values measured in the regridded EBHIS data. Note that these values are significantly higher than the measured values in the VLA data alone. The difference reveals that the galaxy contains a considerable amount of diffuse gas, which cannot be traced by the interferometer. 

It is apparent that within the inner radii the measured flux density for the VLA data is higher than the Effelsberg data. These regions are dominated by small angular scale structures (compared to the EBHIS beam). For all the data sets the measured flux densities for the VLA data reveal a steep drop at larger radii. This is the result of the aforementioned negative bowls around the structure. These regions correspond to large angular scale structures, where the Effelsberg data provide the missing information and compensate this effect.

The results demonstrate that the Effelsberg data can overcome the short-spacing problem and provide the missing short-spacing data. They also show that the naturally weighted data sets lead to better results regarding the short-spacing correction. This is of great important if the focus is on measuring total flux or studying extended structures. 

\section{Summary and outlook} \label{sec:summary}

The new era of radio astronomy will be characterized by new large interferometric arrays. However, the observations of extended Galactic objects as well as many nearby galaxies performed by these new instruments will be subject to short-spacing problem (SSP). This is due to the fact that interferometric arrays are not sensitive to the emission on the largest angular scales, which are important to study the extended and diffuse gas component.

Additionally, data handling is an important aspect for the current, modern and next-generation facilities. Due to the huge amount of raw data produced by such arrays long-term storage of raw data is not feasible. In this paper, a new combination method is introduced to perform the short-spacing correction (SSC) in the image domain. The method operates on reduced, science-ready data. The only inputs are single-dish and interferometric data cubes as FITS files and the corresponding telescope beams. Additional information such as visibilities or dirty beam images are not required. This is a key advantage for the observations of future telescopes such as ASKAP and WSRT/APERTIF as the method can operate \textit{on-the-fly} as part of online data processing tools. The comparison with other methods shows that our approach comes up with very similar results compared those of feathering and combination before deconvolution. Moreover, no Fourier transformation is performed as the method operates in the image space directly. Thus, the resulting combined data product is not subject to aliasing if strong emission is present at the border of the interferometric map. The crucial step in the pipeline is the regridding of the single-dish data, where interpolation inaccuracies can cause flux inconsistency or induce artifacts. However, this can be efficiently circumvented if an appropriate pixel grid is chosen during the single-dish data reduction process such that the difference in the grid resolution of both low- and high-resolution is not large.

We present archival deep \hi observations of the SMC carried out with the 64 m Parkes telescope and the ATCA \citep{1997MNRAS.289..225S, 1999MNRAS.302..417S} and the result of their combination. The result of the combination underlines the importance of the SSC for nearby, extended galaxies with considerable amount of large angular scale structure. It also shows that the combination method meets the expectations regarding the measured flux density and angular resolution.   

Another important consideration is the choice of imaging parameters for interferometric data sets. This topic is of great importance, since re-imaging is not possible if raw data are not stored for future facilities producing large data rates. We study the impact of two imaging parameters, weighting scheme and pixel size, on the reconstructed synthesized image for two nearby galaxies NGC\,4214 and NGC\,5055 from THINGS ensemble \citep{2008AJ....136.2563W}.
Our analysis shows that, as expected, the reconstructed synthesized images from the same raw data can have significantly different properties (e.g., resolution, noise level, sensitivity towards extended structures) depending on the chosen parameters. We also perform the SSC for NGC\,4214. In this case the single-dish data is provided from the Effelsberg-Bonn \hi Survey \citep[EBHIS, ][]{2010ApJS..188..488W, 2011AN....332..637K, 2016A&A...585A..41W}. The results show that for the purpose of the short-spacing correction the natural weighted interferometric data set is the more appropriate choice.

\acknowledgements
The lead author is grateful to the Deutsche Forschungsgemeinschaft (DFG) for support under grant numbers KE757/7-1-3. Frank Bigiel acknowledges support from DFG grant BI1546/1-1. The lead author is very thankful to Tobias R\"ohser and the referee for their very useful comments and suggestions. Based on observations performed by the 64 m Parkes telescope. Based on observation performed by the Australia Telescope Compact Array (ATCA). Based on observation performed by 100-m Effelsberg telescope. Based on observations with the Very Large Array (VLA).

\bibliographystyle{an}
\bibliography{ssc}

\end{document}